\documentclass[prb,aps,amssymb,shownopacs,onecolumn]{revtex4}
\usepackage{amsmath}
\usepackage{amssymb}
\usepackage{amsthm}
\usepackage{amsfonts}
\usepackage{enumerate}
\usepackage{latexsym}
\usepackage{youngtab}

\def\ZZ{{\mathbb{Z}}}
\def\vec#1{{\bf #1}}
\def\oncite#1{\onlinecite{#1}}
\newcommand{\beq}{\begin{equation}}
\newcommand{\eneq}{\end{equation}}

\begin{document}

\tolerance 10000

\newcommand{\vk}{{\bf k}}


\title{Central Charge and Quasihole Scaling Dimensions From Model Wavefunctions: Towards Relating Jack Wavefunctions to W-algebras}

\author{B. Andrei Bernevig}
\affiliation{Princeton Center for Theoretical Science, Princeton, New Jersey 08544, USA}
\affiliation{Department of Physics, Princeton Univeristy, Princeton, New Jersey 08544, USA}
\author{Victor Gurarie}
\affiliation{Department of Physics, CB390, University of Colorado,
Boulder CO, 80309}
\author{Steven H. Simon}
\affiliation{Rudolf Peierls Center for Theoretical Physics, Oxford, OX1 3NP, United Kingdom}

\begin{abstract}
We present a general method to obtain the central charge and quasihole scaling dimension directly from groundstate and quasihole wavefunctions. Our method applies to wavefunctions satisfying specific clustering properties. We then use our method to examine the relation between Jack symmetric functions and certain ${\cal W}$-algebras. We add substantially to the evidence that the $(k,r)$ admissible Jack functions correspond to correlators of the conformal field theory ${\cal W}_k(k+1,k+r)$, by calculating the central charge and scaling dimensions of some of the fields in both cases and showing that they match. For the Jacks described by unitary ${\cal W}$-models, the central charge and quasihole exponents match the ones previously obtained from analyzing the physics of the edge excitations. For the Jacks described by non-unitary ${\cal W}$-models the central charge and quasihole scaling dimensions obtained from the wavefunctions differ from the ones obtained from the edge physics, which instead agree with the ``effective" central charge of the corresponding ${\cal W}$-model.
\end{abstract}

\date{March 3, 2009}

\pacs{72.25.-b, 72.10.-d, 72.15. Gd}

\maketitle

\section{Introduction}

In the lowest Landau level in symmetric gauge\cite{Prange,RMP}, wavefunctions can be thought of as single-valued analytic functions of complex variables.  As a result, many powerful mathematical tools can be brought to bear on the study of lowest Landau level physics.   In particular, the power of conformal field theory\cite{YellowBook} has been useful in understanding fractional quantum Hall wavefunctions.   Starting with the work of Moore and Read\cite{MooreRead} it was realized that correlators of certain conformal field theories  (CFTs) can be used as trial wavefunctions, and further that the wavefunctions would then inherit the nontrivial topological properties of the CFT\cite{MooreRead,RMP}.

Perhaps the most interesting of the quantum Hall states that have been constructed using CFT is the Read-Rezayi series\cite{ReadRezayi} some of which are actually thought to exist in nature\cite{RMP}. These wavefunctions can be described as the densest polynomial wavefunctions that satisfy a particular clustering condition --- that the wavefunction not vanish when $k$ particles come to the same point, but does vanish when the $k+1^{st}$ particle arrives  (this simple rule describes the $\mathbb{Z}_k$ Read-Rezayi wavefunction for bosons, a more complicated rule describes the analogue for fermions).  Because of the success of the Read-Rezayi wavefunctions, generalizations of this clustering rule are worth considering.  Although the rule could be generalized in many different ways, one approach has recently been proposed that seems particularly interesting\cite{Bernevig1,Bernevig2,Bernevig3}.  In this approach quantum Hall wavefunctions are described as being so-called Jack symmetric functions\cite{Jack} (or ``Jacks").  The mathematical structure of the Jacks allows detailed study of these wavefunctions, and the Jacks include the Read-Rezayi wavefunctions, as well as other previously proposed wavefunctions\cite{Gaffnian,Green,WenWu}, as special cases.   Interestingly, the fact that these Jacks obey a generalized clustering rule was previously pointed out in the mathematical literature\cite{Jimbo1,Jimbo2}, and in that work it was conjectured that these Jack polynomials should be describable as correlators of certain ${\cal W}$-algebra CFTs.  This correspondence was proven rigorously in a special case\cite{Jimbo2} (the $k=2$ case, which corresponds to the Virasoro minimal model CFTs $M(3,2+r)$ in a notation we will describe below).  However, for the general case, the connection remains a conjecture. One purpose of this paper is to add substantially to the evidence for this correspondence. We do this by devising a rather general method that can be used to extract the central charge from a wavefunction that exhibits a particular (${\mathbb{Z}}_k$-like) clustering property. The central charge comes out as a coefficient deeply embedded in the ground-state wavefunction. When used on Jacks described by unitary models, our method gives a central charge that equals the one obtained through the fundamentally different method  of counting edge excitations in Ref.~\oncite{Bernevig3}. For non-unitary theories, the edge method and the method derived in this paper result in different values of the central charge.  The results of the edge method\cite{Bernevig3} correspond to the so-called ``effective central charge" of the ${\cal W}$-algebra  whereas the method presented in this paper directly obtains the central charge of the same theory.   We also show how to obtain the fundamental quasihole scaling exponent as a coefficient embedded in the \emph{un-normalized} quasihole wavefunction obtained in Ref.~\oncite{Bernevig3}. When used on Jacks described by unitary models, the scaling dimension appears consistent with the one previously obtained through the computation of edge correlators on the disk in Ref.~\oncite{Bernevig3} (they do not appear consistent for non-unitary models). Although we apply it only to Jack polynomials, our method works for any $k$-clustered wavefunction. The second purpose of this paper is to examine some of the basic properties of the ${\cal W}$-algebras and their applicability as fractional quantum Hall wavefunctions.

The outline of this paper is as follows.  In section \ref{sec:CFTreview} we will briefly review the connection between quantum Hall wavefunctions and conformal field theories, and present the strategy for obtaining the central charge from a generic $k$-clustered wavefunction.  In section \ref{sec:JackIntro} we introduce the Jack polynomial wavefunctions.  We assume that the Jack polynomials are described by a CFT with certain properties and we derive several properties of this putative CFT -- including the central charge in section \ref{sec:jackcentral} and scaling dimensions of certain quasihole fields in section \ref{sec:jackquasi}. In section \ref{sec:Wintro} we introduce the general ${\cal W}$-algebra CFTs (in particular the {$\cal W$}-algebras based on $sl(k)$ or ${\cal W}_k$ algebras).   In section \ref{sec:Wcluster} we discuss construction of quantum Hall wavefunctions using these CFTs.  In section \ref{sec:Wfinal} we show that the CFTs ${\cal W}_{k}(k+1,k+r)$, sometimes notated ${\cal WA}_{k-1}(k+1,k+r)$, precisely match the derived properties of the Jacks.  Rather interestingly we find that, with the exception of the Read-Rezayi series (including the Moore-Read state and the Laughlin states), all of the Jacks correspond to non-unitary CFTs.   General arguments, presented in a series of recent papers by Read\cite{ReadEdge,ReadViscocity} appear to preclude such wavefunctions from representing a gapped phase of matter, although they could correspond to critical points between phases.  Gapless excitations have not yet been identified and the precise meaning of the nonunitarity is still under investigation.   A recent manuscript \cite{jolicoeur} has proposed a method by which unitary, albeit Abelian, theories may be built from nonunitary ones.  This work builds on the observation\cite{Bernevig4,Bernevig5} that the abelian Jain state is a $2$-quasielectron - $2$ quasihole excitation of the non-unitary Gaffnian state (which allows a Jack polynomial description).

We note that other ${\cal W}$-algebras exist which are unitary, although many other CFTs also exist with similar clustering properties. We point out that other $\cal W$-algebra wavefunctions, that would be unitary, are also possible.

\section{Constructing Quantum Hall States Using Conformal Field Theories}
\label{sec:CFTreview}

In this section we review construction of quantum Hall states from CFTs.  More detailed discussion is given in Refs.~\oncite{RMP,MooreRead,ReadRezayi}.   Those familiar with this topic may be able to skip much of this section.

We will consider CFTs with a simple current $\psi_1$ having $\ZZ_k$ symmetry (i.e., it fuses with itself $k$ times to give the identity).  In the original work on parafermions\cite{Zamparafermion}, the operator product expansion (OPE) for such a theory is given generally by the following (with $n = 0, \ldots, k-1$ being defined modulo $k$)
\begin{equation}
\label{eq:psifus}
    \lim_{z \rightarrow z'} \psi_n(z) \psi_{n'}(z') \sim (z -
    z')^{\Delta_{nn'}} \psi_{(n+n'){\rm{mod}}k}(z') + \ldots
\end{equation}
where $\psi_0$ is interpreted as the identity field $I$ and ``$\ldots$"
indicates less singular terms and
\begin{equation}
    \Delta_{nn'} = h_{[(n+n'){\rm{mod}}k]} - h_n  - h_{n'}
\end{equation}
Here $h_n$ is the conformal dimension (scaling dimension, or conformal weight) of the field $\psi_n$ which we will assume is given by the expression
\begin{equation}
\label{eq:dimpsi}
    h_n = \frac{r n(k-n)}{2k}
\end{equation}
with $r \geq 2$ an integer.   The usual parafermions of Fateev and Zamolodchikov\cite{Zamparafermion} are recovered
for $r=2$.  For other values of $r > 2$, we obtain a modified
parafermion-like theory.  Indeed, such a modification was
proposed very briefly in Appendix A of Ref.
\oncite{Zamparafermion}.  We will refer to a CFT of this form as being $(r/2)^{th}$ generation $\ZZ_k$ CFT, and we will use the notation $\ZZ_k^{(r/2)}$ proposed by Ref.~\oncite{Jacob}.     Note that all of the cases of this type that we are aware of with $r$ odd correspond to nonunitary theories.   (The fact that odd  $r$ are allowed was apparently first pointed out in Ref.~\oncite{Ravanini}).  Note that generically, this OPE, without specification of further terms in the expansion, is not sufficient to completely define the CFT, and there may be many allowable CFTs that fit this description of $\ZZ_k^{(r/2)}$.  These different possible theories are distinguished, among other ways, by their central charges.

We will further assume that in the relevant CFT, there are no additional conserved currents. In this case should any two primary fields fuse to give the identity $I$,  conformal invariance gives us\cite{YellowBook}
\begin{equation}
\label{eq:Tfus}
\lim_{z \rightarrow z'}
    \phi(z) \phi'(z')  = (z-z')^{-2h} \left[ I + (z - z')^2 (2 h/c) T(z') + \ldots \right]
\end{equation}
where $h$ is the scaling dimension (conformal weight) of the primary fields $\phi$ and $\phi'$ (these dimensions are necessarily equal if they fuse to $I$), $c$ is the central charge of the theory, and $T$ is the stress-energy tensor that satisfies the OPE\cite{YellowBook}
\begin{equation}
\label{eq:TOPE}
    \lim_{z \rightarrow z'}  T(z) \phi(z')  =  \frac{h}{(z-z')^2} \,  \phi(z') +   \ldots
\end{equation}
for any primary field $\phi$ with scaling dimension $h$.

Given a conformal field theory with these properties, we can construct the multiparticle wavefunction as a correlator\cite{MooreRead,ReadRezayi,RMP}
\begin{equation} \label{eq:Psi}
    \Psi(z_1, \ldots, z_N) =
    \left\langle
    \psi_1(z_1) \psi_1(z_2) \ldots \psi_1(z_N) \right\rangle  \prod_{i < j} (z_i - z_j)^{\frac{r}{k} + M}
\end{equation}
with $M$ a nonnegative integer (throughout most of this paper we will assume $M=0$).   We will assume that the
number $N$ of particles is a multiple of $k$ (otherwise the correlator
term is strictly zero).  Note that the usual Gaussian factors that
occur for wavefunctions in the lowest Landau level are not written
explicitly here (See Refs. \oncite{MooreRead,RMP} for further discussion of this issue).
The case of even $M$ will correspond to a boson
wavefunction and odd $M$ will correspond to a fermion (electron)
wavefunction.

The fusion relation Eq. \ref{eq:psifus} gives us
\begin{equation}
\label{eq:psifus1}
    \lim_{z_1 \rightarrow z_2} \psi_1(z_1) \psi_1(z_2) \sim (z_1 -
    z_2)^{-\frac{r}{k}} \,\, \psi_2(z_2) + \ldots
\end{equation}
This fractional power is precisely canceled by the fractional
Jastrow factor in Eq. \ref{eq:Psi} so that the wavefunction is
properly single valued in the electron coordinate.

It will sometimes be convenient to think of the above Jastrow
factors in Eq. \ref{eq:Psi} as having resulted from vertex
operators $e^{i \beta \varphi(z)}$ for $\varphi$ being a free
massless scalar Bose field satisfying
\begin{equation}
    \langle \varphi(z_1) \varphi(z_2) \rangle = -\log(z_1 - z_2)
\end{equation}
such that we have the operator product expansion
\begin{equation}
    e^{i a \varphi(z_1)} e^{i b \varphi(z_2)} \sim (z_1 -
    z_2)^{ab}
\end{equation}
which results in the conformal weight\cite{YellowBook} (scaling dimension) of $e^{i \varphi \beta}$ being $\beta^2/2$.  Strictly speaking the correlator of these vertex operators is zero unless a neutrality condition is satisfied.  This issue is ignored as we assume a smeared background charge (this background charge also reintroduces the above neglected gaussian factors\cite{MooreRead}).

Now we can define the ``electron" operator
\begin{equation}
    \psi_e(z) = \psi_1(z) e^{i \varphi(z) \beta}
\end{equation}
and choosing
\begin{equation}
\beta =  \sqrt{M + r/k} \end{equation} we can rewrite written
Eq. \ref{eq:Psi} as
\begin{equation}
\label{eq:Psi2}
    \Psi(z_1, \ldots, z_N) =
    \left\langle
    \psi_e(z_1) \psi_e(z_2) \ldots \psi_e(z_N) \right\rangle
\end{equation}
Again for $M$ even this is a fully symmetric wavefunction and for $M$ odd, fully antisymmetric.   By using the OPEs, it is easy to establish that in the $M=0$ case the wavefunction does not vanish as $k$ particles come to the same point, but vanishes as $r$ powers when the $k+1$'st particle arrives:  this is a simple $k$-cluster wavefunction in the notation of Ref. \oncite{SimonProjection}.   As noted in that work, such wavefunctions do not exist for $k r$ odd, and correspondingly no ${\mathbb{Z}}_k^{(r/2)}$ theory exists for $k r$ odd.   For general $M$ the wavefunction vanishes as $n(n-1)M /2$ powers as $n \leq k$ particles come together and vanishes as $(k+1) k M / 2 + r$ powers when the $k+1^{th}$ arrives.

In the case of $M$ even, the wavefunction is fully symmetric corresponding to a wavefunction for bosons and we should expect the elementary ``electron"  field $\psi_e$ to have an integer dimension.  However, with $M$ even the scaling dimension of $\psi_e$ (The sum of the dimensions of $\psi_1$ and the vertex) is integer only for even $r$ and is half-integer for $r$ odd.  Conversely for $M$ odd, one has a fully antisymmetric wavefunction, but $\psi_e$ has half-integer dimension only for $r$ even.   This should make one suspect that there are some problems for the case of odd $r$, and indeed there are no unitary theories for the case of odd $r$ which precludes the possibility of odd $r$ wavefunctions representing gapped phases of matter.  (It would be nice to develop a deeper understanding of precisely how these two facts are related).

Using the arguments of Refs.~\oncite{ReadRezayi,MooreRead,RMP,SimonProjection} it is easy to establish that the degree $N_{\phi}$ of the polynomial wavefunction $\Psi$ is given by
\begin{equation}
    N_{\phi} = \nu^{-1} N - {\cal S}
\end{equation}
where
\begin{equation}
\label{eq:mynu}
    \nu = \frac{k}{r + kM}
\end{equation}
is the filling fraction and
\begin{equation}
    {\cal S} = r + M
\end{equation}
is the so-called ``Shift" on the sphere.   From here on we will be considering quantum Hall effect of bosons and we will consider the case of $M=0$ for simplicity.  Generalization to other values of $M$ is relatively trivial.

The chosen CFT will typically contain many other primary field operators. Suppose the conformal field theory contains a field $\sigma$ with
operator product expansion
\begin{equation} \label{eq:sigfus}
    \lim_{z_1 \rightarrow z_2} \sigma(z_1) \psi_{1}(z_2) \sim (z_1 -
    z_2)^{\Delta_{\sigma 1}} \phi(z_2) + \ldots
\end{equation}
Here we must have
\begin{equation}
\label{eq:sigfusdim}
    \Delta_{\sigma 1} = h_\phi - h_1 - h_\sigma
\end{equation}
where $h_\sigma$ and $h_\phi$ are the conformal weights (scaling dimensions) of the
fields $\sigma$ and $\phi$ respectively.  The fact that there is only one conformal family on the right hand side of Eq.~\ref{eq:sigfus} is guaranteed by the assumption that $\psi_1$ is a simple current.

We then define a quasihole operator
\begin{equation}
    \psi_{qh}^{\sigma}(z) = \sigma(z) e^{i \varphi(z) \gamma}
\end{equation}
where
\begin{equation}
    \gamma = (s - \Delta_{\sigma 1}) / \beta
\end{equation}
with $s$ a nonnegative integer.  This choice of $\gamma$ is the
only possibility that will make the wavefunction $\Psi$ (Eq.
\ref{eq:qh}) properly single valued in the electron coordinates
$z_i$. The resulting wavefunction can be written out as
\begin{eqnarray}
&&     \Psi(z_1, \ldots, z_N; w_1, \ldots w_n)  = \label{eq:qh}
 \left\langle \psi_{qh}(w_1) \ldots \psi_{qh}(w_n)
    \psi_e(z_1) \ldots \psi_e(z_N) \right\rangle \\ &=& \nonumber
 \left\langle \sigma(w_1) \ldots  \sigma(w_n )
    \psi_1(z_1)  \ldots \psi_1(z_N) \right\rangle \\ && \times  \prod_{i < j} (z_i - z_j)^{\beta^2} \prod_{k=1}^N \prod_{m=1}^n (z_k - w_m)^{s-\Delta_{\sigma 1} }
    \prod_{p < m}^n  (w_p - w_m)^{\gamma^2}
          \label{eq:psiwithqps}
\end{eqnarray}
The charge on such a quasihole is given by the exponent  $s - \Delta_{\sigma 1}$ which pushes away a corresponding fraction of
the ambient density from the position of each quasihole.  It is
easy to show\cite{ReadRezayi} that the resulting charge must be
\begin{equation}
\label{eq:qhcharge}
    e^*_{qh}  = \left( s - \Delta_{\sigma 1} \right) e \nu
\end{equation}
where $-e$ is the charge of the electron.   Note that, for example,  by fusing with $p$ electron fields we can also create quasiparticles with change $e^*_{qh} - p e$.

One trivial possibility is to choose the field $\sigma$ to be the
identity field (with dimension zero). In this case, the identity
field fuses with $\psi_1$ to give $\psi_1$ again (so $\phi=\psi_1$
in Eq. \ref{eq:sigfus}) and $\Delta_{\sigma 1} = 0$.  The
quasihole is then given by $s=1$ in (The $s=0$ case gives the
identity operator). We identify this case as the Laughlin
quasihole and we then write
\begin{equation}
\psi_{qh}^{Laughlin}(z) = e^{i \varphi(z)/\beta}
\end{equation}
This can be inserted into a wavefunction resulting in a Jastrow factor.  For example,
\begin{equation}
    \langle \psi_{qh}^{Laughlin}(w) \psi_e(z_1) \ldots \psi_e(z_N) \rangle =  \langle \psi_e(z_1) \ldots \psi_e(z_N) \rangle \,\, \prod_{k=1}^N (w - z_N)
\end{equation}
where the correlator on the right hand side is the wavefunction in the
absence of quasiholes.  The charge of this quasihole is
$e^*_{qh}=e \nu$ as is expected for a quasihole created by a
single flux insertion\cite{Prange}.

More generally, however, for nontrivial fields $\sigma$ the charge
on the quasihole will be some fraction of the Laughlin value.
Furthermore, when several nontrivial (non-Laughlin) quasiholes are
created, the correlator separates into conformal blocks.  This is
the hallmark of nonabelian statistics -- the fact that there are
several orthogonal wavefunctions that describe the set of
quasiholes at one particular set of positions\cite{MooreRead,RMP}.   By examining the
fusion rules of the quasihole operators $\sigma$ we can count the number of conformal blocks and determine the degeneracy associated with the nonabelian statistics.

\subsection{Strategy for obtaining the central charge of a theory from wavefunctions}
\label{sec:strategy}

Given a polynomial wavefunction in the lowest Landau level, we would like to identify a CFT which gives this wavefunction as a correlator.  Let us imagine that we are given a bosonic polynomial wavefunction $\Psi(z_1,z_2,z_3,z_4,\ldots,z_N)$ with several properties that make it compatible with CFTs of parafermion type as described above.  We assume that the wavefunction does not vanish as $k$ particles come together, but vanishes as $r$ powers when the $k+1$st arrives (we are assuming $M=0$ bosons in the language above).  We further assume that it is a simple cluster wavefunction\cite{SimonProjection}, meaning that it is filling fraction $\nu=k/r$ with shift $r$.     As described above, such wavefunctions are compatable with CFTs of parafermion type, and in particular can be compatible with $\ZZ_k^{(r/2)}$ CFTs.

We extract the putative correlator  by removing the Jastrow factor (See Eq.~\ref{eq:Psi})
\begin{equation}
\phi_0(z_1,z_2, \ldots ,z_N)=\frac{\Psi(z_1,z_2, \ldots ,z_N)}{\prod_{i<j}^N (z_i-z_j)^{1/\nu}}
\end{equation}
We will further assume that the CFT we are searching for is rational --- that the number of primary fields is finite.   Our general strategy will be to successively fuse $k$ coordinates together to obtain the identity field again (See Eq.~\ref{eq:psifus}).  The subleading term of the final fusion (See Eq. \ref{eq:Tfus}) allows us to produce the stress-energy tensor $T$ with a coefficient that depends on the central charge.

Now let $z_1 =z_2 + \epsilon_1$ and let $\epsilon_1 \rightarrow 0$; then expand in $\epsilon_1$.  Given the expected OPE Eq.~\ref{eq:psifus} we should obtain something of the form
\begin{equation}
\phi_0(z_2+\epsilon_1,z_2,\ldots,z_N) = \frac{1}{\epsilon_1^{2h_1 - h_2}} (\phi_1(z_2,\ldots,z_N) +\epsilon_1 \phi_{1,1}(z_2,\ldots,z_N) +  \epsilon_1^2 \phi_{1,2}(z_2,\ldots,z_N) + \ldots) \label{fusion1}
\end{equation}
Since the function $\Psi$ is given to us, we can easily determine $\phi_1, \phi_{1,1}, \ldots $ explicitly. If the divergence exponent is not $2 h_1 - h_2$ then we conclude that the CFT (if it exists) is not of the $\ZZ_k^{(r/2)}$ parafermion type.  In terms of correlators, the function $\phi_1$ should be given by
\begin{equation}
    \phi_1 =
    \langle \psi_2(z_2) \psi_1(z_3) \psi_1(z_4) \ldots \psi_1(z_N) \rangle
\end{equation}

Now assuming that $k > 2$ we repeat the procedure, taking $z_2 = z_3+ \epsilon_2$ and let $\epsilon_2 \rightarrow 0$; then expand in $\epsilon_2$ to obtain
\begin{equation}
\phi_1(z_3+\epsilon_2,z_3, \ldots ,z_N) = \frac{1}{\epsilon_2^{h_1 + h_2- h_3}} (\phi_2(z_3, \ldots ,z_N) +\epsilon_2 \phi_{2,1}(z_3,\ldots,z_N) + \epsilon_2^2 \phi_{2,2}(z_3,\ldots,z_N)+ \ldots)
\end{equation}
We continue this procedure $k-2$ times.  We finally obtain
\begin{equation}
    \phi_{k-2} = \langle \psi_{k-1}(z_{k-1}) \psi_1(z_k) \psi_1(z_{k+1}) \ldots \psi_1(z_N) \rangle
\end{equation}
Taking the last limit obtains
\begin{equation}
\label{eq:last}
\phi_{k-2}(z_{k}+\epsilon_{k-1},z_k, \ldots ,z_N) = \frac{1}{\epsilon_{k-1}^{2 h_1}} (\phi_{k-1}(z_k, z_{k+1},  \ldots ,z_N) + \epsilon_{k-1}^2 \phi_{k-1,2}(z_k,z_{k+1},\ldots,z_N)+ \ldots)
\end{equation}
Here we have used the OPE Eq.~\ref{eq:Tfus}.
\begin{equation}
    \phi_{k-1}(z_k, z_{k+1},  \ldots ,z_N)  = \langle I(z_k) \psi_1(z_{k+1}) \ldots \psi_1(z_N) \rangle = \langle  \psi_1(z_{k+1}) \ldots \psi_1(z_N) \rangle
\end{equation}
which should be independent of the position $z_{k}$.    Note there is no term $\phi_{k-1,1}$ linear in $\epsilon_{k-1}$ in the expansion (again, if this is not true, it is evidence that we do not have a CFT of parafermion type).  Indeed, it will be useful below to note that the subleading term vanishes when the leading term is the identity.   The second term, on the other hand, from Eq.~\ref{eq:Tfus}, gives us
\begin{equation}
    \phi_{k-1,2}(z_k, z_{k+1},  \ldots ,z_N)  = (2 h_1/c) \langle T(z_k) \psi_1(z_{k+1}) \ldots \psi_1(z_N) \rangle
\end{equation}
We now take one more limit setting $z_k = z_{k+1}+ \epsilon_k$ and let $\epsilon_k \rightarrow 0$; then expand in $\epsilon_k$.  Using the OPE Eq.~\ref{eq:TOPE} we obtain
\begin{equation}
\label{eq:iso}
    \phi_{k-1,2}(z_{k+1} + \epsilon , z_{k+1},  \ldots ,z_N)  = \frac{1}{\epsilon_{k}^{2}} \left[  \phi_{k}(z_{k+1}, \ldots, z_N)  + \ldots \right]
\end{equation}
where
\begin{equation}
\phi_k = (2 h_1^2/c) \langle  \psi_1(z_{k+1}) \ldots \psi_1(z_N) \rangle
\end{equation}
Thus by taking the ratio
\begin{equation}
\label{eq:rat}
 \frac{\phi_k(z_k, \ldots, z_N) }{ \phi_{k-1}(z_k, \ldots, z_N)} =  2 h_1^2/c
\end{equation}
we are able to extract the putative central charge. A similar scheme will also be used below to extract scaling dimensions of quasiparticles.

\section{Jack Wavefunctions}

\subsection{The Basics of using Jack Symmetric Functions as Quantum Hall Wavefunctions}
\label{sec:JackIntro}

In this section we describe the construction of quantum Hall wavefunctions as Jack symmetric functions.   This reviews work of Refs.~\oncite{Bernevig1,Bernevig2,Bernevig3,Jimbo1,Jimbo2,Stanley}.

The Jack symmetric functions (Jacks) are polynomials satisfying a number of particular properties.  We refer the reader to Ref.~\oncite{Jack,Stanley,MacDonald,Demmel,Lassalle} for a more detailed discussion of many of these properties.   We write a general Jack as
\begin{equation}
J^{\alpha}_{\lambda}(z_1,z_2, \ldots, z_N)
\end{equation}
This is a function of $N$ complex variables $z_i$, and parametrically depends on a real so-called ``Jack-parameter" $\alpha$, as well as a  partition $\lambda$ of length $|\lambda|$ where $|\lambda| \leq N$.   A partition $\lambda$ is an ordered set of numbers $\lambda_i \leq \lambda_{i-1} $, $1 \le i \le \left| \lambda \right|$ such that
\begin{equation}
    \sum_{i=1}^{\left| \lambda \right|} \lambda_i = \ell_\lambda,
\end{equation}
where $\ell_\lambda$ is some integer number.  Each partition can be uniquely associated with a Young diagram\cite{MacDonald} in the standard way.  Note that we have followed the usual convention that a partition is made up of positive integers with no integer equal to zero.  However, frequently below we will want to think of the partition $\lambda$ as having exactly $N$ pieces, thus we can do this by including in addition $N - |\lambda|$ occurrences of the integer 0.

A detailed definition of the Jack polynomial is given in appendix \ref{app:defjack} for the interested reader.   For the present, however, it suffices to state that the Jacks are simply polynomials satisfying a great number of interesting properties that have been previously worked out\cite{Jack,Stanley,MacDonald,Demmel,Lassalle}.

In recent work by Haldane and Bernevig\cite{Bernevig1,Bernevig2,Bernevig3} it was pointed out that setting the Jack parameter
\begin{equation}
\label{eq:alpha}
\alpha = -(k+1)/(r-1)
 \end{equation}
with $k+1$ and $r-1$ coprime, generates symmetric polynomials (which we think of as bosonic quantum Hall wavefunctions, corresponding to $M=0$ above) satisfying the admissibility condition that the wavefunction vanishes as $r$ powers when $k+1$ particles come to the same point.  This admissibility condition had been noted previously in the mathematical literature\cite{Jimbo1,Jimbo2} and for the Read-Rezayi states by Haldane \cite{HaldaneAPS}. In Ref.~\oncite{Bernevig1} it was shown that the requirement of translational invariance uniquely selects \emph{all} the Jacks that can be good FQH wavefunctions, as we will discuss further below. Translational invariance immediately gives the Jack parameter $\alpha=-(k+1)/(r-1)$ as well as the $(k,r)$ admissibility on partitions.   Among the wavefunctions that can be described as Jacks are the Read-Rezayi series\cite{ReadRezayi} (including the Moore-Read state\cite{MooreRead}), the Laughlin wavefunctions\cite{Prange}, and the Gaffnian\cite{Gaffnian}.    We emphasize that the Jack parameters here are negative rational, in contrast with other applications to condensed matter systems such as the Calogero model\cite{Calogero}, which have positive Jack parameter.

An important ingredient of Jack polynomials is the so-called root state. This can be constructed out of the partition $\lambda$ according to
\begin{equation} R_\lambda ={\cal S} \left[  \prod_{i=1}^N z_i^{\lambda_i} \right]/{\cal N},
\end{equation}
where the symbol ${\cal S}$ represents the symmetrization over all the permutations of $z_i$ (equivalently, one may think of it as computing the permanent perm$(z_i^{\lambda_j})$), and ${\cal N}$ is the normalization factor which we define below in (\ref{normm}).   It is convenient to think of a root state as representing occupation of orbitals.   Imagining the orbitals $\varphi_m \sim z^m$ in the lowest Landau level in the plane, we describe the root state $R_\lambda$ as an set of occupation numbers $n_m(\lambda)$ for bosons occupying orbitals where summing the total number of particles in all orbitals gives the total number of particles $N$.    In other words, $n_m(\lambda)$ is the number of times the integer $m$ occurs in the partition $\lambda$. For example,
a root polynomial corresponding to the partition $\lambda_1=2$, $\lambda_2=0$, given by $R_\lambda = z_1^2 +z_2^2$, can be described in terms of occupation numbers $n_0=1$, $n_1=0$,  $n_2=1$.
In this paper, we will almost always write partitions in terms of occupation numbers.  For example, for the root polynomial introduced in this paragraph we
will write $\lambda=[1,0,1]$, where the terms in the square brackets are $n_m(\lambda)$ for $m=0,1,2$. (See Refs. \oncite{Bernevig1} and \oncite{Bernevig2} for more details on the translation between the orbital occupation representation of partitions and the conventional representation of partitions). In terms of $n_m(\lambda)$ the normalization factor ${\cal N}$ is given by
\begin{equation} \label{normm} {\cal N} = \prod_m n_m!.
\end{equation}
Its role is simply to eliminate any additional factors which might arise from symmetrizing already symmetric expressions.

A Jack symmetric polynomial is not  simply equal to its root state.  Each Jack is labeled by a root state, but is actually a superposition of the root state along with many other descendant states which can be constructed by ``squeezing" occupation numbers --- i.e., which can be obtained by starting with the root state and moving bosons towards each other in pairs\cite{Bernevig1,Bernevig2}.   In the language of partitions the descendent states are {\it dominated} by the root, or highest weight, state.  The Jack is given by a particular combination of the root state and its descendants which make it an eigenvalue of a differential operator known as the Laplace-Beltrami operator\cite{Stanley}.

Not every Jack polynomial can correspond to a quantum Hall wave function.  Indeed, some of them are not even translationally invariant (that is, change under the
change of variables $z_i \rightarrow z_i +a$).   In Ref.~\oncite{Bernevig1} it was shown that the Jacks that
correspond to translationally invariant wavefunctions are (a) Those with Jack parameter as in Eq.~\ref{eq:alpha}  and (b) have partitions corresponding to root states with the property that no more than $k$ bosons may occupy $r$ consecutive orbitals, for some given $k$ and $r$. It is interesting that in the limit of a thin cylinder, the root state is precisely the wavefunction, which means that the entire physics just becomes an issue of distributing bosons so as to satisfy the admissibility condition. This fact has been exploited in a number of recent publications\cite{Bergholtz,Bergholtz2,Seidel,ArdonneTorus,ArdonneTorus2}.

Given $N$ coordinates, with $N$ divisible by $k$, the root partition that yields the lowest degree polynomial (and hence the highest density wavefunction) is given by the occupation numbers
\begin{equation}
    \lambda = [k\underbrace{0 \,\, 0\,\, \ldots \,0}_{\mbox{\small $r-1$ times}}k\underbrace{0 \,\, 0\,\, \ldots \,0}_{\mbox{\small $r-1$ times}} k \ldots \underbrace{0 \,\, 0\,\, \ldots \,0}_{\mbox{\small $r-1$ times}} k]
\end{equation}
where there are $N/k$ orbitals filled with $k$ bosons each.   We abbreviate this occupation with the obvious notation
\begin{equation}
    \lambda = [k \, 0^{r-1} \, k \, 0^{r-1} \, k \ldots 0^{r-1} \, k]
\end{equation}
If we consider the corresponding Jack symmetric function to be a wavefunction for bosons
\begin{equation}
\label{eq:Jackwavefunction}
 \Psi = J^\alpha_{k \, 0^{r-1} \, k \, 0^{r-1} \, k \ldots 0^{r-1} \, k}(z_1, \ldots, z_N)
\end{equation}
with $\alpha$ as above, we generate a wavefunction that vanishes as $r$ powers when $k+1$ coordinates approach each other, but does not vanish when $k$ coordinates come to the same point.   As discussed in Refs. \oncite{Bernevig1,Bernevig2} this is a wavefunction at filling fraction $\nu=k/r$ with shift ${\cal S}=r$.  Thus this is a simple $k$-cluster wavefunction as discussed in Ref.~\oncite{SimonProjection}.  Further, this suggests that such a wavefunction may be described as a correlator of a $\ZZ_k^{(r/2)}$ CFT as described in the previous section (with $M=0$).   We will show additional evidence below that this is indeed the case.

One can similarly describe quasihole states in terms of Jacks that have lower density root partitions.  For example, if we allow the $N$ bosons to occupy one additional orbital, admissible root partitions include
\begin{align}
 [0 k \, 0^{r-1} \, k &0^{r-1} \, k \ldots 0^{r-1} \, k]  \\
 [1 (k-1) \, 0^{r-1} \, k& 0^{r-1} \, k \ldots 0^{r-1} \, k]  \\
 [k \, 0^{r} \, k&0^{r-1} \, k \ldots 0^{r-1} \, k]  \\
[k \, 0^{r-1} \, (k-2)  2& 0^{r-1} \, k \ldots 0^{r-1} \, k]  \\
& \vdots \nonumber
\end{align}
and many others.  These many possibilities correspond both to the multiple types of quasiparticles as well as the many positions where the quasiparticles may be placed.   One may analyze all of the possibilities to categorize the possible quasiparticle types, and further one can consider how these quasiparticle types fuse with each other. This exercise has been performed\cite{ArdonneTorus,ArdonneTorus2} in the context of the thin cylinder limit and it has been found that these admissibility rules correspond to the particles and fusion rules of $su(k)_r$.

\subsection{Central Charge of the Jack Polynomials}
\label{sec:jackcentral}

We now extract the central charge for the Jack polynomial wavefunctions.  We consider a wavefunction corresponding to the $(k,r)$ Jack as in Eq.~\ref{eq:Jackwavefunction}, fixing $\alpha$ as in Eq.~\ref{eq:alpha} throughout (we do not write the parameter $\alpha$ explicitly from here on).

We now proceed as in section \ref{sec:strategy}  to extract the central charge.   As above, we begin with
\begin{equation}
\phi_0 =
\langle \psi_1(z_1) \psi_1(z_2)\ldots\psi_1(z_N) \rangle = \frac{J_{k0^{r-1}k0^{r-1}k\ldots k0^{r-1}k}}{\prod_{i<j=1}^N (z_i - z_j)^\frac{1}{\nu}}
\end{equation}
Without loss of generality, since the polynomials are translationally invariant\cite{Bernevig1}, we may choose $z_1=0$.    This is quite convenient as the only orbital which does not vanish as $z \rightarrow 0$ is the $m=0$ orbital, so it is easy to see when certain wavefunctions do or do not vanish.   Further, the remaining Jack polynomial, having taken this limit is given by the partition $[(k-1)0^{r-1}k0^{r-1}k\ldots k0^{r-1}k]$ with a coefficient of unity since all our Jack polynomials share the ``monic" normalization.

Now take the limit $z_2 \rightarrow 0$, and and expand as in Eq.~\ref{fusion1}.  For $k>2$ there will be a term proportional to  $z_2$ in the expansion (like in Eq.~\ref{fusion1}), which means we have not fused to the identity, and we simply take the leading term of the expansion, which is given by the Jack with partition $[(k-2)0^{r-1}k0^{r-1}k\ldots k0^{r-1}k]$. In the process, from the divergence of the Jastrow factor we find the relation:
\begin{equation}
h_1 + h_1 - h_2 = \frac{1}{\nu}
\end{equation} We now take the coordinate $z_3$ to zero, and so forth, continuing like this until we reach $J_{10^{r-1}k0^{r-1}k\ldots k0^{r-1}k}$. During this process, keeping track of successive divergencies of the Jastrow factor we find the series of equations for the scaling dimensions which are:
\begin{equation}
h_1 + h_m - h_{m+1} =\frac{m}{\nu}
\end{equation}
where $m=1 \ldots k-1$ (with $h_k=0$), which is consistent with our expectation from Eq.~\ref{eq:dimpsi}.

We now focus on the last fusion of this series (See Eq.~\ref{eq:last}) from which we will get the central charge.  Here, we let $z_{k} \rightarrow 0$:
\begin{eqnarray} \nonumber
 & & \frac{J_{10^{r-1}k0^{r-1}k\ldots k0^{r-1}k}(z_{k},z_{k+1},\ldots,z_N)}{\prod_{i=k}^N z_i^{\frac{k-1}{\nu}} \prod_{i<j= k}^N (z_i-z_j)^\frac{1}{\nu}} =
 \\ & & \frac{1}{z_k^{\frac{k-1}{\nu}} \prod_{i=k+1}^N z_i^{\frac{k}{\nu}} \prod_{i<j= k+1}^N (z_i-z_j)^\frac{1}{\nu}}\left( J_{0^{r}k0^{r-1}k\ldots k0^{r-1}k}(z_{k+1},\ldots,z_N) + z_k P_1 + z_k^2 P_2 + \ldots \right) \nonumber \\ & & \times  \left(1+ z_k \frac{1}{\nu} \sum_{j=k+1}^N \frac{1}{z_j} + z_k^2(\frac{1}{2}\frac{1}{\nu}(1+ \frac{1}{\nu}) \sum_{j=k+1}^N \frac{1}{z_j^2} + \frac{1}{\nu^2} \sum_{i<j =k+1}^N \frac{1}{z_i z_j}) + \ldots \right) \label{electronfusion}
\end{eqnarray}
The polynomials $P_1, P_2$ can be obtained by expanding the Jack polynomial $J_{10^{r-1}k0^{r-1}k\ldots k0^{r-1}k}(z_{k},z_{k+1},\ldots,z_N)$ for small $z_k$.  In this expansion, the resulting polynomials are in fact other Jacks with the same value of $\alpha$, i.e., they are $(k,r)$ admissible.  This expansion is shown explicitly in appendix \ref{app:expansion} giving
\begin{eqnarray}
\nonumber
& & J_{10^{r-1}k0^{r-1}k\ldots k0^{r-1}k}(z_{k},z_{k+1},\ldots,z_N) = J_{0^{r}k0^{r-1}k\ldots k0^{r-1}k}(z_{k+1},\ldots,z_N) \\ &+&  z_{k} A_1 J_{0^{r-1}1(k-1)0^{r-1}k\ldots k0^{r-1}k}(z_{k+1},\ldots,z_N)  \nonumber + z_k^2\left[ B_1 J_{0^{r-2}10(k-1)0^{r-1}k\ldots k0^{r-1}k}(z_{k+1},\ldots,z_N) \right. \\ & + &  \left.B_2 J_{0^{r-1}1(k-1)0^{r-2}1(k-1)0^{r-1}k\ldots k0^{r-1}k}(z_{k+1},\ldots,z_N) \right] + \ldots  \label{eq:expand}
\end{eqnarray}
\noindent
For simplicity of notation, let us define the following notation:
\begin{eqnarray}
&P_0 = J_{0^{r}k0^{r-1}k\ldots k0^{r-1}k}(z_{k+1},\ldots,z_N) \nonumber \\ & P_{1,0} = J_{0^{r-1}1(k-1)0^{r-1}k\ldots k0^{r-1}k}(z_{k+1},\ldots,z_N)  \nonumber \\ & P_{2,0} = J_{0^{r-2}10(k-1)0^{r-1}k\ldots k0^{r-1}k}(z_{k+1},\ldots,z_N) \nonumber \\ & P_{2,1} = J_{0^{r-1}1(k-1)0^{r-2}1(k-1)0^{r-1}k\ldots k0^{r-1}k}(z_{k+1},\ldots,z_N)
\end{eqnarray}
where, of course, in Eq.~\ref{electronfusion} we have $P_1 = A_1 P_{1,0}$ and $P_2= B_1 P_{2,0} + B_2 P_{2,1}$.  The coefficients $A_1, B_1, B_2$ are to be determined. Of these coefficients, $A_1, B_1$ are simple to evaluate whereas $B_2$ is somewhat harder. Fortunately, we will not actually need to fully evaluate $B_2$. Leaving the values of these coefficients unspecified for the moment (we derive them in appendix \ref{app:expansion}), we have, for the products of brackets in Eq.~\ref{electronfusion}:
\begin{eqnarray}
&\frac{1}{z_k^{\frac{k-1}{\nu}} \prod_{i=k+1}^N z_i^{\frac{k}{\nu}} \prod_{i<j= k+1}^N (z_i-z_j)^\frac{1}{\nu}}\left( J_{0^{r}k0^{r-1}k\ldots k0^{r-1}k}(z_{k+1},\ldots,z_N) + z_k P_1 + z_k^2 P_2  + \ldots \right) \times  \\ & \times  \left(1+ z_k \frac{1}{\nu} \sum_{j=k+1}^N \frac{1}{z_j} + z_k^2(\frac{1}{2}\frac{1}{\nu}(1+ \frac{1}{\nu}) \sum_{j=k+1}^N \frac{1}{z_j^2} + \frac{1}{\nu^2} \sum_{i<j =k+1}^N \frac{1}{z_i z_j})  + \ldots \right)  = \nonumber \\  & = \frac{1}{z_k^{\frac{k-1}{\nu}} \prod_{i=k+1}^N z_i^{\frac{k}{\nu}} \prod_{i<j= k+1}^N (z_i-z_j)^\frac{1}{\nu}}( P_0 + z_k (A_1 P_{1,0}  +   \frac{1}{\nu} \sum_{j=k+1}^N \frac{1}{z_j} \cdot P_0) +  \nonumber  \\ & z_k^2\left( (\frac{1}{2}\frac{1}{\nu}(1+ \frac{1}{\nu}) \sum_{j=k+1}^N \frac{1}{z_j^2} + \frac{1}{\nu^2} \sum_{i<j =k+1}^N \frac{1}{z_i z_j})\cdot P_0 + A_1 P_{1,0} \cdot \frac{1}{\nu} \sum_{j=k+1}^N \frac{1}{z_j} + B_1 P_{2,0} + B_2 P_{2,1} \right) + \ldots ) \nonumber
\end{eqnarray}
\noindent
The coefficient $A_1$  is derived in appendix \ref{app:expansion} and is given by $A_1 = -\frac{1}{\nu}$.   We also have, as shown in appendix \ref{app:onefurther} the identity
\begin{equation}
\sum_i \frac{\partial}{\partial z_i} P_0 = r P_{1,0}
\end{equation}
We can thus derive a series of identities:
\begin{eqnarray}
& \sum_{i}\frac{\partial}{\partial z_i} J_{0^{r}k0^{r-1}k\ldots k0^{r-1}k}(z_{k+1},\ldots,z_N)  =  r J_{0^{r-1}1(k-1)0^{r-1}k\ldots k0^{r-1}k}(z_{k+1},\ldots,z_N) = \nonumber \\ & = \sum_{i}\frac{\partial}{\partial z_i} \prod_j z_j^r J_{k0^{r-1}k\ldots k0^{r-1}k}(z_{k+1},\ldots,z_N) = r \sum_i \frac{1}{z_i} \prod_j z_j^r  J_{k0^{r-1}k\ldots k0^{r-1}k}(z_{k+1},\ldots,z_N) = \nonumber \\ &=r \sum_i \frac{1}{z_i}J_{0^{r}k0^{r-1}k\ldots k0^{r-1}k}
\end{eqnarray}
where we have used the fact that $J_{k0^{r-1}k\ldots k0^{r-1}k}(z_{k+1},\ldots,z_N)$ is a highest weight state by virtue of being a groundstate. The above shows that:
\begin{equation}
P_{1,0} = \sum_i \frac{1}{z_i} P_0
\end{equation}  Then we have
\begin{equation}
(A_1 P_{1,0}  +   \frac{1}{\nu} \sum_{j=k+1}^N \frac{1}{z_j} \cdot P_0) = \frac{1}{\nu} (-P_{1,0}  +  \sum_{j=k+1}^N \frac{1}{z_j} \cdot P_0) =0
\end{equation} so the first order term in $z_k$ vanishes as it should. We now have:
\begin{eqnarray}
& \frac{1}{\prod_{i=k+1}^N z_i^{\frac{k}{\nu}} \prod_{i<j= k+1}^N (z_i-z_j)^\frac{1}{\nu}}\left( J_{0^{r}k0^{r-1}k\ldots k0^{r-1}k}(z_{k+1},\ldots,z_N) + z_k P_1 + z_k^2 P_2 \right) \times \\ & \times  \left(1+ z_k \frac{1}{\nu} \sum_{j=k+1}^N \frac{1}{z_j} + z_k^2(\frac{1}{2}\frac{1}{\nu}(1+ \frac{1}{\nu}) \sum_{j=k+1}^N \frac{1}{z_j^2} + \frac{1}{\nu^2} \sum_{i<j =k+1}^N \frac{1}{z_i z_j})\right)  = \nonumber \\  & = \frac{1}{ \prod_{i=k+1}^N z_i^{\frac{k}{\nu}} \prod_{i<j= k+1}^N (z_i-z_j)^\frac{1}{\nu}}( P_0 +z_k^2\left( (\frac{1}{2}\frac{1}{\nu}(1 - \frac{1}{\nu}) \sum_{j=k+1}^N \frac{1}{z_j^2} - \frac{1}{\nu^2} \sum_{i<j =k+1}^N \frac{1}{z_i z_j})\cdot P_0  + B_1 P_{2,0} + B_2 P_{2,1} \right) + \ldots) \nonumber
\end{eqnarray}
Note that $k/\nu = r$.  A few additional identities, shown in appendix \ref{app:onefurther}, are now also useful
\begin{equation}
\label{eq:efirst}
\frac{P_0}{\prod_{i=k+1} z_i^{k/\nu}}= J_{k0^{r-1}k\ldots k0^{r-1}k}(z_{k+1},\ldots,z_N)
\end{equation}
\begin{equation}
\label{eq:esecond}
\frac{P_{2,0}}{\prod_{i=k+1} z_i^{k/\nu}}= \frac{1}{\prod_{i=k+1} z_i^{2}} J_{10(k-1)0^{r-1}k\ldots k0^{r-1}k}(z_{k+1},\ldots,z_N)
\end{equation}
\begin{equation}
\label{eq:ethird}
\frac{P_{2,1}}{\prod_{i=k+1} z_i^{k/\nu}}= \frac{1}{\prod_{i=k+1} z_i} J_{1(k-1)0^{r-2}1(k-1)\ldots k0^{r-1}k}(z_{k+1},\ldots,z_N)
\end{equation}
Now as in Eq.~\ref{eq:iso} let $z_{k+1} \rightarrow 0$, and isolate the singularities in $1/z_{k+1}^2$.   We only need the leading singularity, so we can immediately see that $P_{2,1}$ doesn't matter because it is less singular (after being multiplied by $\prod_{i=k+1} z_i^{-k/\nu}$). Also when $z_{k+1} \rightarrow 0$ we have, to leading order:
\begin{eqnarray}
& \frac{1}{\prod_{i=k+1} z_i^{k/\nu}} (\frac{1}{2}\frac{1}{\nu}(1 - \frac{1}{\nu}) \sum_{j=k+1}^N \frac{1}{z_j^2} - \frac{1}{\nu^2} \sum_{i<j =k+1}^N \frac{1}{z_i z_j})\cdot P_0 = \frac{1}{2}\frac{1}{\nu}(1 - \frac{1}{\nu})  \frac{1}{z_{k+1}^2}J_{k0^{r-1}k\ldots k0^{r-1}k}(z_{k+1}=0,\ldots,z_N) = \nonumber \\ & = \frac{1}{2}\frac{1}{\nu}(1 - \frac{1}{\nu})  \frac{1}{z_{k+1}^2}J_{(k-1)0^{r-1}k\ldots k0^{r-1}k}(z_{k+2},\ldots,z_N)
\end{eqnarray}
also,
\begin{eqnarray} \nonumber
\frac{P_{2,0}}{\prod_{i=k+1} z_i^{k/\nu}} &=&  \frac{1}{\prod_{i=k+1} z_i^{2}} J_{10(k-1)0^{r-1}k\ldots k0^{r-1}k}(z_{k+1},\ldots,z_N)  \\ & \underbrace{=}_{z_{k+1} \rightarrow 0}&  \frac{1}{z_{k+1}^2} \frac{1}{\prod_{i=k+2} z_i^{2}} J_{10(k-1)0^{r-1}k\ldots k0^{r-1}k}(z_{k+1}=0,\ldots,z_N) \nonumber \\ &=& \frac{1}{z_{k+1}^2} \frac{1}{\prod_{i=k+2} z_i^{2}} J_{00(k-1)0^{r-1}k\ldots k0^{r-1}k}(z_{k+2},\ldots,z_N) \nonumber \\&=& \frac{1}{z_{k+1}^2} J_{(k-1)0^{r-1}k\ldots k0^{r-1}k}(z_{k+2},\ldots,z_N)
\end{eqnarray}
We are now in a position to obtain the final equation: as $z_{k+1} \rightarrow 0$ we have:
\begin{eqnarray}
& \frac{1}{ \prod_{i=k+1}^N z_i^{\frac{k}{\nu}}}( P_0 +z_k^2\left( (\frac{1}{2}\frac{1}{\nu}(1 - \frac{1}{\nu}) \sum_{j=k+1}^N \frac{1}{z_j^2} - \frac{1}{\nu^2} \sum_{i<j =k+1}^N \frac{1}{z_i z_j})\cdot P_0  + B_1 P_{2,0} + B_2 P_{2,1} \right))= \nonumber \\ &= J_{k0^{r-1}k\ldots k0^{r-1}k}(z_{k+1},\ldots,z_N) + \frac{1}{z_{k+1}^2} ( \frac{1}{2}\frac{1}{\nu}(1 - \frac{1}{\nu}) + B_1) J_{k-10^{r-1}k\ldots k0^{r-1}k}(z_{k+2},\ldots,z_N))
\end{eqnarray}
The coefficient $B_1$ is derived in appendix \ref{app:expansion} giving
\begin{equation}
B_1 = \frac{r(r-1)}{2}\frac{\alpha+1}{((r-1)\alpha +1)((r-2)\alpha +1)}= -\frac{r}{2 k} \frac{(r-k-2)(r-1)}{- k r + 2 k +1}
\end{equation}
Taking the appropriate ratio as in Eq.~\ref{eq:rat}, we then have
 \begin{equation}
 \frac{2 h_1^2}{c} = \frac{1}{2}\frac{1}{\nu}(1 - \frac{1}{\nu}) + B_1
 \end{equation} where $h_1 = \frac{r(k-1)}{2 k}$.  With trivial algebra we obtain
 \begin{equation}
 \label{eq:central1}
c = r\frac{k-1}{k+r} (2 k +1 - k r)
 \end{equation}
This is precisely the central charge of the ${\cal W}_{k}(k+1,k+r)$ CFT\cite{Fateev1,Fateev2,SchoutensReview} as will be discussed further below.

\subsection{Quasihole Scaling Exponent for Jack Polynomials}
\label{sec:jackquasi}

In this section, we obtain the quasihole scaling dimension from the polynomial wavefunctions.   The strategy is quite similar to that described above --- fusion of fields together to form the stress energy tensor as in Eq.~\ref{eq:Tfus} giving a prefactor proportional to the field dimension.

We must first write down a wavefunction that describes an elementary quasihole at some position $w$ and another object with which it can uniquely fuse to form the identity at the origin.  This other object is precisely the fusion of $k-1$ quasiholes.   These wavefunctions can be established by invoking the clustering conditions as described in Ref.~\oncite{Bernevig3}
\begin{eqnarray}
& &\left.   \Psi(w ; z_1, \ldots, z_N)\right|_{z_1 = z_2 \ldots = z_k = w} = 0 \\
& & \left.  \Psi(w ; z_1, \ldots, z_N)\right|_{z_1 = z_2=0} = 0
\end{eqnarray}
The unique such admissible wavefunction is a superposition of the Jacks given by\cite{Bernevig3}
\begin{equation}
\Psi(w;z_1,\ldots,z_N) = \sum_{i=0}^{N/k} \left(-\frac{w}{k} \right)^i |i \rangle
\end{equation}
where $|i \rangle$ are the Jack polynomials
\begin{eqnarray}
    |0 \rangle &=& J_{0 k 0^{r-1} k 0^{r-1} k \ldots 0^{r-1} k 0^{r-1} k} \\
    |1 \rangle &=& J_{1 (k-1) 0^{r-1} k 0^{r-1} k \ldots 0^{r-1} k 0^{r-1} k} \\
    |2 \rangle &=& J_{1 (k-1) 0^{r-2} (k-1) 0^{r-1} k \ldots 0^{r-1} k 0^{r-1} k} \\
    &\vdots&
\end{eqnarray}
and the Jack parameter $\alpha$ is as above always taken to be $-(k+1)/(r-1)$.

If we had generated this wavefunction from a conformal field theory (See Eq.~\ref{eq:qh}) there would generically be an additional nonsingle valued dependence on the quasihole coordinate $w$  (i.e., there may be branch cuts).    However, this really just provides a normalization for the wavefunction whereas the Jacks are not normalized with respect to integration over all the $z$ coordinates (and supposedly the CFT correlator is normalized, see Ref.~\oncite{Nayak,Gurarie,ReadViscocity} for detailed discussion of this issue).  We thus expect that there may be some arbitrary prefactor $f(w,0)$ which multiplies our Jack wavefunction in order to produce the CFT wavefunction.

We thus write a proposed correlator corresponding to a single quasihole at $w$ and $k-1$ quasiholes at the origin.
\begin{equation}
\langle \sigma'(0) \sigma(w) \psi_e(z_1)\psi_e(z_2) \ldots \psi_e(z_N) \rangle = f(w,0) \frac{\Psi (w;z_1,\ldots,z_N)}{\prod_{i=1}^N (z_i - 0)^{\frac{k-1}{k}} (z_i- w)^{\frac{1}{k}}}
\end{equation}
Here we have notated the primary field associated with the quasihole as $\sigma$ and the field associated with the fusion of $k-1$ quasiholes as $\sigma'$.  These two fields fuse to give the identity as in Eq.~\ref{eq:Tfus}.  We now want to let the quasihole coordinate $w \rightarrow 0$ and examine the result of this fusion. Keeping terms up to order  $w^2$ we obtain
\begin{eqnarray}
&\langle \sigma'(0) \sigma(w) \psi_e(z_1) \ldots \psi_e(z_N) \rangle = \frac{f(w,0)}{\prod_{i=1}^N z_i}\left(|0\rangle - \frac{w}{k} |1\rangle + \left(\frac{w}{k}\right)^2 |2\rangle  + \ldots \right) \times \nonumber \\ & \times \left(1+ \frac{1}{k} w\sum_{i=0}^N \frac{1}{z_i} + w^2 \left(\frac{1}{2}\frac{1}{k}(1+ \frac{1}{k}) \sum_{i=0}^N \frac{1}{z_i^2} + \frac{1}{k^2} \sum_{i<j} \frac{1}{z_i z_j}\right) + \ldots \right)
\end{eqnarray}
We now massage the terms in the product of the two brackets. The term independent of $w$ is $|0 \rangle$.  Thus, the unknown prefactor $f(w,0)$ must contain the divergent prefactor shown in Eq.~\ref{eq:Tfus}.   The term linear in $w$ is
\begin{equation}
 \frac{1}{k}\left( - |1 \rangle +  \sum_{i=0}^N \frac{1}{z_i} |0\rangle\right)
\end{equation}
Considering the form of Eq.~\ref{eq:Tfus} we now must show that this linear term vanishes.   This is demonstrated explicitly in appendix \ref{app:vanish}.  We now go to the second order terms and write them as:
\begin{equation}
\left(\frac{1}{2}\frac{1}{k}(1+ \frac{1}{k}) \sum_{i=0}^N \frac{1}{z_i^2} + \frac{1}{k^2} \sum_{i<j} \frac{1}{z_i z_j}\right) |0\rangle - \frac{1}{k^2}\sum_{i=0}^N \frac{1}{z_i} |1\rangle + \frac{1}{k^2} |2\rangle
 = \left(\frac{1}{2}\frac{1}{k}(1- \frac{1}{k}) \sum_{i=0}^N \frac{1}{z_i^2} - \frac{1}{k^2} \sum_{i<j} \frac{1}{z_i z_j}\right) |0\rangle + \frac{1}{k^2} |2\rangle \end{equation}
\noindent where we have again used the identity $\sum_i (1/z_i) |0\rangle = |1\rangle$ proved in appendix \ref{app:vanish}. So we have hence proved that, for $w \rightarrow 0$ we have:
\begin{equation}
\langle \sigma'(0) \sigma(w) \psi(z_1) \ldots \psi(z_N) \rangle = \frac{f(w,0)}{\prod_{i=1}^N z_i}\left( |0\rangle + w^2 (\left(\frac{1}{2}\frac{1}{k}(1- \frac{1}{k}) \sum_{i=0}^N \frac{1}{z_i^2} - \frac{1}{k^2} \sum_{i<j} \frac{1}{z_i z_j}\right) |0\rangle + \frac{1}{k^2} |2\rangle)  + \ldots  \right)
\end{equation}
\noindent To make this look nicer, we can write (again using the same identity)
$|0\rangle = J_{0k0^{r-1}k0^{r-1}k\ldots k0^{r-1}k} = \prod_{i=1}^N z_i J_{k0^{r-1}k0^{r-1}k\ldots k0^{r-1}k} = \prod_{i=1}^N z_i \Psi_{GS}$ where $\Psi_{GS}$ is the ground state in the absence of the quasiholes (Eq.~\ref{eq:Psi2}).  Hence we have
\begin{equation}
\langle \sigma'(0) \sigma(w) \psi_e(z_1) \ldots \psi_e(z_N) \rangle = f(w,0)\left( \Psi_{GS} + w^2 (\left(\frac{1}{2}\frac{1}{k}(1- \frac{1}{k}) \sum_{i=0}^N \frac{1}{z_i^2} - \frac{1}{k^2} \sum_{i<j} \frac{1}{z_i z_j}\right) \Psi_{GS} + \frac{1}{k^2}  \frac{1}{\prod_{i=1}^N z_i} |2\rangle) + \ldots \right)
\end{equation}
Thus from Eq.~\ref{eq:Tfus} we have
\begin{equation}
(2 h_\sigma/c) \langle T(0) \psi_e(z_1) \ldots \psi_e(z_N) \rangle = \left(\frac{1}{2}\frac{1}{k}(1- \frac{1}{k}) \sum_{i=0}^N \frac{1}{z_i^2} - \frac{1}{k^2} \sum_{i<j} \frac{1}{z_i z_j}\right) \Psi_{GS} + \frac{1}{k^2}  \frac{1}{\prod_{i=1}^N z_i} |2\rangle
\end{equation}
Note that $T$ is the stress energy tensor of the CFT not including the bose vertex operators.  We now let one of the electron coordinates approach the position $0$ giving a leading singularity
\begin{equation}
\lim_{z_1 \rightarrow 0} (2 h_\sigma/c) \langle T(0) \psi_e(z_1) \ldots \psi_e(z_N) \rangle = \frac{1}{z_1^2} \frac{1}{2}\frac{1}{k}(1- \frac{1}{k}) \Psi_{GS} + \ldots
\end{equation}
From the OPE Eq.~\ref{eq:TOPE} we thus conclude that
\begin{equation}
2 \frac{h_{\sigma} h_1}{c} = \frac{1}{2}\frac{k-1}{k^2}
\end{equation}
Using the above expression (Eq.~\ref{eq:dimpsi}) for $h_1$ we then obtain
\begin{equation}
\label{eq:hres}
h_{\sigma} = \frac{c}{2 k r}
\end{equation}
thus relating the scaling dimension of the quasiparticle to the central charge.   As expected this result gives precisely the scaling dimension of the minimal quasiparticle field of the ${\cal W}_{k}(k+1,k+r)$ CFT\cite{Fateev1,Fateev2,SchoutensReview} as will be discussed further below.

In fact, since we also know that the charge of the quasiparticle is $-e \nu/k$ (there is a $k$-fold fractionalization of the Laughlin quasiparticle), then by Eqs.~\ref{eq:sigfusdim} and \ref{eq:qhcharge} we know immediately that the fusion of the quasiparticle with the electron field must create another field $\phi$ with scaling dimension
\begin{equation}
\label{eq:dimvar}
h_{\phi}  = h_1 + h_{\sigma} + 1/k
\end{equation}
Indeed, again in the ${\cal W}_{k}(k+1,k+r)$ CFT, the fusion of $\sigma$ with $\psi_1$ creates a field of precisely this scaling dimension, as we will see below.

In principle, with enough effort, one can calculate the scaling dimension of other fields using similar techniques.

\section{${\cal W}$-algebra Quantum Hall Wavefunctions}
\label{sec:Walgebra}

In this section we describe a very large family of CFTs
known as ${\cal W}$-algebras.  As discussed above these CFTs may be used in construction of quantum Hall wavefunctions, and further, some of them apparently correspond to the wavefunctions obtained from Jack polynomials.

\subsection{Introductory Facts about some ${\cal W}$-algebras}
\label{sec:Wintro}

A great deal is known about ${\cal
W}$-algebras, and the variety of ${\cal W}$-algebras that have
been studied in the literature is both vast an growing.  We refer
the reader to Ref. \oncite{SchoutensReview} for general
information about this field. In the current paper we will focus
on some of the simplest cases known which are the minimal models
of the ${\cal W}_{k}$ algebras.  These minimal models were
first described in Refs. \oncite{Bais,Fateev1}. In this paper
we will follow the approaches of Refs.
\oncite{Fateev1,Fateev2}, and then specify (and simplify) to
the situations in which we are interested  (There are, however, other ways to describe the same CFTs, See Ref.~\oncite{SchoutensReview}.   Nothing in this
section is new, but is rather just a distillation of prior results
from these references.

Each of the simple Lie algebra ${\cal A}_{n-1}, {\cal B}_n, {\cal
D}_n, {\cal E}_6, {\cal E}_7, {\cal E}_8$ can be associated with a
${\cal W}$-algebra. Among these we will only be interested in
${\cal WA}_{k-1}$ which has $\mathbb{Z}_k$ symmetry.   These are sometimes notated as just
${\cal W}_{k}$.   We leave for
future research the question of what quantum Hall states can be
constructed from other ${\cal W}$-algebras\cite{SchoutensReview}.

Focusing on the algebra ${\cal W}_k$, a field  $\Phi(\vec l;
\vec l')$ is specified by two vectors $\vec l$ and $\vec l'$ of
$k-1$ positive integers
\begin{eqnarray}
    (\vec l;\vec l') = (l_1, \ldots, l_{k-1} \, ; \, \, l'_1, \ldots l'_{k-1})
\end{eqnarray}
The fusion of such fields can be written symbolically as
\begin{equation}
    \Phi(\vec l; \vec l') \times \Phi(\vec m; \vec m') = \sum_{(\vec s; \vec
    s')}\left[ \Phi(\vec s; \vec s')  \right]
\end{equation}
The values of $(\vec s; \vec s')$ that contribute to the right
hand side for a given $(\vec l; \vec l')$ and $(\vec m; \vec m')$
is known as the ``qualitative structure" or ``selection rules" of
the algebra.  Treating $(\vec l; \vec l')$ as the corresponding
specified representation of $sl((k) \times sl(k)$ gives the proper
structure.  Equivalently, we think of the vectors
\begin{eqnarray}
& &    (l_1 - 1, \, l_2 -1, \, \ldots, \, l_{k-1} -1) \\
 &&    (l'_{k-1} - 1,\,  \ldots, \, l'_2 -1, \, l'_1 -1)
\end{eqnarray}
as the Dynkin labels of two $SU(k)$
representations\cite{YellowBook} so that $(\vec l; \vec l')$ are
labels for a representation of $SU(k) \times SU(k)$.   The
qualitative structure of the fusion in $SU(k) \times SU(k)$ then
is the qualitative structure of our ${\cal W}$-algebra.  It is
useful to note that when all of the $l_i$ values are unity, we
have the trivial, or identity, representation

To relate these labels to more familiar notation, we note that
Dynkin labels can be trivially converted to Young
tableau\cite{YellowBook}.   In $SU(k)$, the $k-1$ labels give the
respective differences between the number of boxes in each of the
$k-1$ successive rows.  So for example, for $k=5$ the Dynkin labels
$(3,0,2,0)$ represents the Tableau
\begin{equation}
\raisebox{-5pt}{$\Yboxdim.1in
    \young(\hfil\hfil\hfil\hfil\hfil,\hfil\hfil,\hfil\hfil)$}
\end{equation}
The identity representation is the empty tableau (although for
clarity we will sometimes write $\bf 1$). Note that to write a field
configuration, we must increment each Dynkin label by one, and note that the  primed indices are read right to left and the unprimed left to right.  Thus,
a typical field for the
case of $k=5$ we might write graphically as
\begin{equation}
    \Phi(4,1,3,1 \, ; \, 2,2,2,1) = \left(\raisebox{-5pt}{$\Yboxdim.1in
    \young(\hfil\hfil\hfil\hfil\hfil,\hfil\hfil,\hfil\hfil)$} \,\, ; \,\, \raisebox{-8pt}{$\Yboxdim.1in
    \young(\hfil\hfil\hfil,\hfil\hfil\hfil,\hfil\hfil,\hfil)$}  \,  \right)
\end{equation}
A minimal model has parameters tuned so that a
finite algebra closes.  In particular, we specify a minimal model
by two integers $p$ and $p'$ both greater than $k$ which
are relatively prime.  This theory is typically called ${\cal W}_k(p,p')$ or ${\cal WA}_{k-1}(p,p')$.  The
fields $\Phi(\vec l; \vec l')$ that form a closed algebra are
given by the set satisfying the constraint
\begin{eqnarray}
    \sum_{i=1}^{k-1} l_i &\leq& p' - 1 \\
    \sum_{i=1}^{k-1} l_i' &\leq& p - 1 \label{eq:res1}
\end{eqnarray}
These constraints can be thought of as restricting the
corresponding Young tableaus to have no more than $p' - k$
and $p - k$ columns respectively.  This type of restriction
in the number of columns of a tableau is analogous to what happens
when one looks at representations of $SU(k)$ at level $m$ where
$m$ is $p' - k$ or $p - k$ respectively.

Furthermore, in the minimal model there is an association of
fields
\begin{equation}
\label{eq:ident}
    \Phi(\vec l; \vec l') =  \Phi(\tilde {\vec l}_m; \tilde {\vec
    l'}_m) ~~~~~~~ m=1, \ldots,  (k-1)
\end{equation}
where
\begin{eqnarray}
   \tilde {\vec l}_m &=& (l_{k-m+1}, \ldots, l_{k-1}, l_0, l_1, \ldots,
   l_{k-m-1})\\
    \tilde {\vec l}'_m &=& (l_{k-m+1}', \ldots, l'_{k-1}, l'_0, l'_1, \ldots, l'_{k-m-1})
\end{eqnarray}
with
\begin{eqnarray}
    l_0 &=& p' - \sum_{i=1}^{k-1} l_i  \\
    l'_0 &=& p - \sum_{i=1}^{k-1} l'_i
\end{eqnarray}

The number of Young tableau in $SU(k)$ with no more than $m$
columns (and as usual no more than $k-1$ rows) is given by
${k-1}\choose m$. Thus, we can apparently write down $(p -1)!
(p' -1)! / ((p - k)! (p' - k)! (k-1)!^2)$  fields
$\Phi$, specified by a combination of two different Young tableaus
(one with no more than $p - g$ rows and one with no more than
$p' - g$ rows).  However, the $g$-fold identification of
fields in Eq. \ref{eq:ident} leaves us with only
\begin{equation}
\frac{(p -1)! \, (p' -1)!}{ (p - k)! \,(p' -
k)! \, (k-1) \, !k!}
\end{equation}
distinct primary fields in this algebra.

It is convenient to define\cite{endnotealpha}
\begin{eqnarray}
    \alpha_{\pm} &=& \frac{\pm 1}{\sqrt{2}} \left(p/ p'
    \right)^{\pm \frac{1}{2}} \\
    \alpha_0 &=& \alpha_+ + \alpha_- = (p -
    p') / \sqrt{2  p p'}
\end{eqnarray}
The central charge of this algebra is
\begin{eqnarray}
    c_k(p;p') &=& (k-1)\left(1 - 2 k (k+1) \alpha_0^2 \right) \\
              &=& (k-1) \left(1 - \frac{k (k+1) ( p - p')^2}{p p'} \right) \label{eq:central}
\end{eqnarray}
The algebra is unitary\cite{Mizoguchi} if and only if  $p = p' \pm 1$, which is the only
case where the central charge can be positive.   In a dynamical theory, the central charge has the interpretation of a density of states or heat capacity.

The scaling dimension (or conformal weight) of a field $\Phi(\vec l; \vec l')$ is given by
\begin{eqnarray}
\label{eq:dimeq}
 & &    h(\vec l; \vec l') = -\alpha_0^2  k (k^2 - 1)/12 +   \\ & & \nonumber \left[
\sum_{i=1}^{k-1}
    \sum_{j=1}^{k-1}  (l_i \alpha_+ + l_i' \alpha_-) F_{ij}  (l_j \alpha_+ + l_j'
    \alpha_-)\right]
\end{eqnarray}
Where $F$ is a symmetric $k-1$ dimensional matrix $F$ with matrix
elements
\begin{equation}
  F_{ij} = j (k-i)/k     ~~~~~ j \leq i
\end{equation}
(This is the inverse of the Cartan matrix of $SU(k)$).

This theory has a conserved $\mathbb{Z}_k$ charge.  The charge of
a field $\Phi(\vec l ;  \vec l')$ is given by
\begin{equation}
\label{eq:qeq}
    q = \left( \sum_{n=1}^{k-1} \left[ m n (l_n -1) - m' n (l_n'-1) \right]
    \right)
    \mod k
\end{equation}
where $m =  p \mod k$ and $m' = p' \mod k$.

Finally we note that for nonunitary theories, an interesting quantity to define is the so-called ``effective central charge" given by
\begin{equation}
    \tilde c = c - 24 h_{min}
\end{equation}
where $h_{min}$ is the conformal weight of the primary field with the smallest (most negative) dimension (sometimes known as the pseudovacuum field).    This quantity is necessarily positive, and like the central charge in unitary theories, represents a density of levels.  For the minimal ${\cal W}_k(p,p')$ theories being considered in this section we have\cite{Dunning}
\begin{equation}
\label{eq:ceff}
    \tilde c_k(p,p') =     (k-1) \left(1 - \frac{k (k+1) }{p p'} \right)
\end{equation}

\subsection{Cluster Wavefunctions from {$\cal W$}-algebras}
\label{sec:Wcluster}

Examining the structure of this ${\cal W}$-algebra, we note that
there always exist simple currents corresponding to the desired parafermion
fields from Eq. \ref{eq:psifus}.  Given $p$ and $p'$
relatively prime and greater than $k$, we examine the minimal
model ${\cal W}_{k}(p, p')$.  In this algebra, we
find simple currents
\begin{eqnarray}
    \psi_1 &=& \Phi(p' - k+1, 1, 1, \ldots, 1 \, ; \, 1, 1, \ldots,
    1)  \\
   \psi_2 &=& \Phi(1, p' - k+1, 1,  \ldots, 1 \, ; \, 1, 1,
   \ldots,
    1)  \\
     & \vdots &  \nonumber \\
    \psi_{k-1} &=& \Phi(1, 1, \ldots, 1, p' - k+1 \, ; \, 1, 1,
    \ldots,
    1)
\end{eqnarray}
and of course the identity operator is given by
\begin{equation}
    \psi_0 =  \Phi(1, 1, 1, \ldots, 1 \, ;  \, 1, 1, \ldots,
    1)
\end{equation}
It is worth noting that due to the field identification Eq.
\ref{eq:ident} these parafermion fields can equally well be
expressed as
\begin{eqnarray}
    \psi_1 &=& \Phi(1, 1, \ldots, 1 \, ; \, 1, 1, \ldots, 1,
    p - k + 1)  \\
   \psi_2 &=& \Phi(1, 1, \ldots, 1 \, ; \,  1, 1,
   \ldots,
    p - k + 1, 1)  \\
     & \vdots &  \nonumber \\
    \psi_{k-1} &=& \Phi(1, 1, \ldots, 1 \, ; \, p - k+1, 1,
    \ldots,
    1)
\end{eqnarray}
In terms of Young tableau we can express these fields compactly
(in both representations)  as
\begin{eqnarray}
    \psi_n &=& \left( \raisebox{8pt}{$\begin{array}{l}{p'-k \,\, \mbox{columns}} \\
     \overbrace{  {\Yboxdim.1in
    {\young(\hfil\hfil\hfil\hfil\hfil,\hfil\hfil\hfil\hfil\hfil,\hfil\hfil\hfil\hfil\hfil)}}
     } \raisebox{10pt}{$\left. \rule[-2.1pt]{0pt}{15pt} \right\}$ {$n$ \,
     \mbox{rows}}}
    \end{array}$} \,\,; \,\,  {\bf 1} \right)  \\
&=& \left({\bf 1}  \,\,; \,\,   \raisebox{8pt}{$\begin{array}{l}{p-k \,\, \mbox{columns}} \\
     \overbrace{  {\Yboxdim.1in
    {\young(\hfil\hfil\hfil\hfil\hfil,\hfil\hfil\hfil\hfil\hfil,\hfil\hfil\hfil\hfil\hfil)}}
     } \raisebox{10pt}{$\left. \rule[-2.1pt]{0pt}{15pt} \right\}$ {$n$ \,
     \mbox{rows}}}
    \end{array}$} \right)
\end{eqnarray}
where we have written $\bf 1$ for the identity representation, or
the empty tableau.

Using Eq. \ref{eq:dimeq} we determine that the scaling dimension of these
parafermion fields is indeed given by the desired Eq. \ref{eq:dimpsi} with
\begin{equation}
    r = (p - k)(p'-k)
\end{equation}
Hence ${\cal
W}_{k}(p, p')$ is a CFT of $\mathbb{Z}_k^{(r/2)}$ type as described above.  Using Eq. \ref{eq:Psi} we find that the algebra ${\cal
W}_{k}(p, p')$ constructs a quantum Hall wavefunction
at filling fraction and shift (See Eq. \ref{eq:mynu})
\begin{equation}
\nu = \frac{k}{(p-k)(p' -k) -  k M}  ~~~~~~~~~~~~~~~~~ {\cal S} = (p-k)(p'-k) + M
\end{equation}
for $M$ a nonnegative integer.   For $M=0$ this wavefunction has the property that it is a simple $k$-cluster wavefunction, that is it does not vanish when $k$ particles come to the same point, but it vanishes as $r$ powers when the $k+1^{st}$ arrives.  We emphasize that for arbitrary $r$, there are generically many CFTs that can correspondingly generate many inequivalent wavefunctions with this property (it appears however, that at least for $r=2$, and $k=2,r=3$, the CFT is uniquely defined by this property).  Note that the unitary CFTs correspond to $p=p' \pm 1$.   Thus we expect a series of unitary wavefunctions at filling fraction and shift
\begin{equation}
\nu = \frac{k}{m(m+1) -  k M}  ~~~~~~~~~~~~~~~~~ {\cal S} = m(m+1) + M
\end{equation}
with $m > k$.  Note again that $m=2$  is the Read-Rezayi series.  The case of $m=3$ and $k=2$ corresponds to the tricritical Ising CFT, which has previously been proposed for a quantum Hall wavefunction by Refs.~\oncite{Schweigert,ReadViscocity}.

\subsection{The series ${\cal W}_{k}(k+1,k +r)$}
\label{sec:Wfinal}

The ${\cal W}$-algebras of interest corresponding to the above discussed Jack polynomials are ${\cal W}_{k}(k+1,k +r)$.  Plugging $p=k+1$ and $p'=k+r$ into the above expression gives us a wavefunction with the properties described in the above section.     Here, we must choose $k$ and $r$ both integers $\geq 2$ and where $k+1$ and $k+r$
are relatively prime.  As discussed above, these algebras have $\mathbb{Z}_k$ symmetry,
and central charge (plugging into Eq.~\ref{eq:central})
\begin{equation}
    c =  \frac{(k-1)(1 - k(r-2)) r}{k+r}
\end{equation}
which matches the central charge of the Jack polynomial found in Eq.~\ref{eq:central1}.

The case of $r=2$ is the $\mathbb{Z}_k$
parafermion\cite{Zamparafermion} CFT model which describes the
corresponding $\mathbb{Z}_k$ Read-Rezayi wavefunctions\cite{ReadRezayi}.
It is easy to see that the $r=2$ case is the only value of $r$ for
which the central charge is positive, and is hence the only case
where ${\cal W}_{k}(k+1,k +r)$ is unitary. The case of $k=2$ simplifies to precisely
the Virasoro minimal model\cite{YellowBook} ${\cal M}(3,2+r)$. The
$k=2, ~ r=2$ case here, which is the the Ising conformal field
theory, corresponds to the Moore-Read Pfaffian\cite{MooreRead}.
The $k=2,~ r=3$ case corresponds to the recently discussed
Gaffnian wavefunction\cite{Gaffnian}.

In addition to the $\psi_n$ fields described above, we now examine a possible quasiparticle field.
Let us consider a field
\begin{equation}
    \sigma =\left( \Yboxdim.1in
    {\young(\hfil)} \, ; \, {\bf 1} \right)
\end{equation}
whose $\mathbb{Z}_k$ charge is $1$ and scaling dimension is (see Eq.~\ref{eq:qeq} and \ref{eq:dimeq})
\begin{equation}
    h_\sigma=\frac{(k-1)(1 + k ( 2 - r))}{2k (r+k)}
\end{equation}
It is easy to check that this indeed satisfies the predicted relationship Eq.~\ref{eq:hres}.
The fusion of the simple current with this elementary spin field is
particularly simple.    We have
    \begin{equation}
        \psi_1 \times \sigma = \phi
    \end{equation}
where $\phi$ is the field
\begin{equation}
    \phi =\left( \raisebox{8pt}{$\begin{array}{c}{r \,\, \mbox{columns}} \\
     \overbrace{  {\Yboxdim.1in
    {\young(\hfil\hfil\hfil\hfil\hfil,\hfil)}}
     }
    \end{array}$}  \,\, ;  {\bf 1} \right)
\end{equation}
which has scaling dimension (again using Eq.~\ref{eq:dimeq})
\begin{equation}
h_{\phi} = \frac{2 k^2 - (k+r)^2 + k (r^2 -3)}{2 k (r+k)}
\end{equation}
In Eq. \ref{eq:sigfus} we have
\begin{equation}
\label{eq:spinfusdim}
    \Delta_{\sigma 1} = h_\varphi - h_\sigma - h_1 = -1/k
\end{equation}
as predicted by Eq.~\ref{eq:dimvar} above.  Thus, the quantum Hall state generated by this CFT has a quasiparticle of charge $e^* = -e \nu/k$.  It is easy to check that this is the lowest charge quasiparticle that can be constructed from the theory.

From Eq.~\ref{eq:ceff} we calculate the effective central charge $\tilde c$ which we find to be given by
\begin{equation}
 \tilde c =  \frac{r(k-1)}{k+r}
\end{equation}
Interestingly this is the value of central charge found in Ref.~\oncite{Bernevig3} by counting the density of edge modes on a disk for $(k,r)$ Jack wavefunctions.  (Strictly speaking, this reference finds $\tilde c+1$, where the $+1$ corresponds to the $U(1)$ charge boson).   Thus it appears to be the effective central charge that determines the density of states.  This is perhaps not surprising since the central charge of the nonunitary ${\cal W}$-models is negative, and a negative density of states would be unphysical.   Also in Ref.~\oncite{Bernevig3} the scaling dimension of the quasiparticle is bounded numerically.  While the result of this calculation agrees with the ${\cal W}$-model prediction for unitary cases, it does not agree for the nonunitary cases.  This apparently contradictory result is not currently understood.

\section{Summary}

We have shown here that the $(k,r)$ Jack quantum Hall states likely correspond to the ${\cal W}_k(k+1,k+r)$ CFTs.   We have shown that both have simple currents $\psi_n$ with ${\mathbb{Z}}_k$ symmetry having scaling dimension given by Eq.~\ref{eq:dimpsi}.  They both have an elementary quasiparticle field $\sigma$ having scaling dimension given by Eq.~\ref{eq:hres}, and $\psi_1$ fuses with $\sigma$ to yield another field $\varphi$ with scaling dimension Eq.~\ref{eq:dimvar}.   While this does not completely prove that the two theories are equivalent, it is very strong evidence.  We comment that for the case of $k=2$ a full proof of equivalence has been given by Ref.~\oncite{Jimbo2}.

Making connection to prior work of Ref.~\oncite{Bernevig3} we find that the central charge determined in that work by edge state analysis in that work agrees with the ${\cal W}$-algebra central charge in the unitary cases, and corresponds to the effective central charge of the ${\cal W}$-algebra in the nonunitary cases.  However, the analysis of the quasiparticle exponent in that work does not appear to be in agreement with the current ${\cal W}$-algebra analysis in nonunitary cases.

It is interesting that amongst all of the wavefunctions described by the Jack polynomials, only the Read-Rezayi series and the Laughlin series correspond to unitary CFTs.  Presumably this means\cite{ReadViscocity,ReadEdge} that, other than these specific cases, the Jack polynomials cannot correspond to gapped phases of matter.  Nonetheless, they can still correspond to critical points between phases, and understanding the nature of this criticality can teach us much about the adjacent phases.  Other unitary ${\cal W}$-algebras could in principle correspond to gappped phases.  We note however, as mentioned in Ref.~\oncite{ReadViscocity}, that identification of a CFT for use as a quantum Hall wavefunction does not yet imply that this wavefunction is the ground state of a Hamiltonian.  Further work will be required to try to construct such Hamiltonians\cite{Simon3} for any proposed wavefunction.

\vspace*{10pt}

{\it Acknowledgements:} B.A.B thanks F.D.M. Haldane for countless discussions and suggestions.   B.A.B. also wishes to thank P. Bonderson, N. Regnault, L. Tevlin, R. Thomale and E. Ardonne for discussions.  V.G. acknowledges support from NSF grant, DMR-0449521.   V.G. also acknowledges helpful discussions with C. Nayak.
S.H.S acknowledges helpful discussions with P. Fendley,  E. Ardonne,
   J. Slingerland, and particularly E. Rezayi and N. Read.

\appendix

\section{Detailed Definition of the Jack Polynomials}
\label{app:defjack}

There are several equivalent ways to define the Jack polynomials\cite{Stanley,MacDonald,Jack,Demmel,Lassalle}.  The reader should note, however, that there are several standard inequivalent normalizations that are used.  Note also, that in this appendix the standard description of a partition is used rather than the occupation basis description.

Let us write the form of the Jack polynomial from Ref.\oncite{Demmel} (see also Ref.\oncite{Stanley}).
We start by defining the Jacks which are a function of a single variable
\begin{equation}
        J_{k}^{\alpha }(z_1)=z_1^k(1+\alpha)\cdots (1+(k-1)\alpha)
\end{equation}
We then define each Jack in terms of Jacks with one fewer variables
\begin{equation}
J^\alpha_{\lambda} (z_1, z_2,...,z_N) = \sum_{\mu \subseteq \lambda} J^\alpha_{\mu} (z_2, ...,z_N) z_1^{|\lambda/\mu|} \beta_{\lambda \mu}
\end{equation}
the summation is over all subpartitions $\mu$ of $\lambda$ ($\mu \subseteq \lambda$) such that the skew partition $\lambda/\mu$
is a  so-called horizontal strip. Here, $\lambda/\mu$ is a horizontal strip if $\lambda_1\ge \mu_1 \ge \lambda_2 \ge \mu_2 \ge \lambda_3 \ge \mu_3...$, or in other words, if no two distinct points of $\lambda/\mu$, regarded as the difference $\lambda -\mu$ of their Young diagrams, lie in the same column (draw the diagram of $\lambda$ then color only the squares of $\lambda$ that don't belong to $\mu$; if no two colored squares are in the same column then $\lambda/\mu$ is a horizontal strip). The exponent $|\lambda/\mu|$ in the above equation is just the difference $\sum_i \lambda_i  - \sum_j \mu_j$ (Note that $\lambda$ and $\mu$ do not have to have identical length).

The coefficients $\beta_{\lambda,\mu}$ are given by (See proposition 6.1 of Ref.~\oncite{Demmel}):
\begin{equation}
\beta_{\lambda, \mu} = \frac{\prod_{(i,j) \in \lambda} B^\lambda_{\lambda, \mu}(i,j)}{\prod_{(i,j) \in \mu} B^\mu_{\lambda, \mu}(i,j)}
\end{equation}
where
\begin{equation}
B^\nu_{\lambda, \mu}(i,j) = h^*_\nu (i,j)  \;\;\; \text{if} \;\;\; \lambda_j' =\mu_j' ; \;\;\;  = h^\nu_* (i,j)  \;\;\; \text{otherwise}
\end{equation}
In the above $\lambda'$ and $\mu'$ are the conjugate partitions of $\lambda$, $\mu$, obtained by transposing the Young diagram of the partition $\lambda$, $\mu$ (i.e. one writes them as Young diagrams and then interchanges the rows with the columns). $h^*_\lambda$ and $h^\lambda_*$ are the generalized upper and lower hook lengths of the partition $\lambda$:
\begin{equation}
h^*_\lambda (i,j) = \lambda'_j-i + \alpha (\lambda_i - j +1), \;\;\;\; h_*^\lambda (i,j) = \lambda'_j-i +1+ \alpha (\lambda_i - j )
\end{equation}
the product $\prod_{(i,j) \in \lambda}$ means product over ALL the pairs $(i,j)$ in the Young tableau of partition $\lambda$, i.e. $i$ goes over all the components of the partition $\lambda=(\lambda_i)$ whereas, for a set $i$, $j$ runs from $1$ to $\lambda_i$.

Unfortunately, this relatively simple definition is not the normalization that we use within the current paper.   The above normalization corresponds to the normalization of Stanley\cite{Stanley}, in which the coefficient of the root monomial of the Jack polynomial $J^\alpha_\lambda$ is equal to $v_{\lambda \lambda} (\alpha)$ where
\begin{equation}
v_{\lambda \lambda} (\alpha) =\prod_{(i,j) \in \lambda} h^\lambda_*(ij)
\end{equation}
In the current paper, on the other hand, we use the ``monic" normalization where which the coefficient of the ``root" (or dominating) monomial is equal to $1$. Hence the coefficients $\beta_{\lambda \mu}$ in our case read:
\begin{equation}
\beta_{\lambda, \mu} = \frac{v_{\mu\mu}(\alpha)}{v_{\lambda \lambda}(\alpha)} \frac{\prod_{(i,j) \in \lambda} B^\lambda_{\lambda, \mu}(i,j)}{\prod_{(i,j) \in \mu} B^\mu_{\lambda, \mu}(i,j)}
\end{equation}
Note that there are cancelations between the $v$'s and the $B$'s which lead to a simplified form, which we do not write explicitly.

\section{Expansion of the Jack Polynomials}
\label{app:expansion}

In the main text Eq. \ref{eq:expand} we need to find the expansion of the Jack of partition $\lambda =  [10^{r-1}k 0^{r-1} k ... k 0^{r-1}k]$ into Jacks of one fewer variable.   Using the recursive definition of the Jack functions in the above appendix, this expansion is straightforward, and we can easily obtain the coefficients of the Jacks corresponding to the partitions
\begin{eqnarray}
\mu^{A_1} &=& [0^{r-1}1k-1 0^{r-1} k ... k 0^{r-1}k] \\
\mu^{B_1} &=& [0^{r-2}10k-1 0^{r-1} k ... k 0^{r-1}k] \\
\mu_{B_2} &=& [0^{r-1}1k-1 0^{r-2} 1k-1 ... k 0^{r-1}k]
\end{eqnarray}
Note that since the coefficient $B_2$ is not needed for the calculation of the central charge, we will not derive it here since it is tedious.

\subsection{Coefficient $A_1$}

Let us now compute the coefficient $A_1= \beta_{\lambda \mu^{A_1}}$. For the two partitions $\lambda$ and $\mu^{A_1}$ we have $\lambda_i = \mu^{A_1}_i$ $\forall$ $i = 1,..., N-k-1$ where $N$ is the original number of particles ($N= l(\lambda)+ k = l(\mu^{A_1}) + k$, where $l(\lambda)$ is the length of the partition $\lambda$ - i.e. the number of elements of $\lambda$ without counting the zeroes -- which means the length of $\lambda$ is also equal to the length of $\mu^{A_1}$). The only place where $\lambda$ differs from $\mu^{A_1}$ is the last element $\mu^{A_1}_{N-k} = \lambda_{N-k} -1= r-1$. For the conjugate partitions $\mu^{A_1'} $ and $\lambda'$ we again have $\mu^{A_1'}_i  =\lambda_i'$ $\forall$ $i=1,...,r-1$ and $i=r+1,..., N_\Phi = \frac{r}{k}(N-k)$. The only place where they differ is $\mu^{A_1'}_r  = \lambda_r' -1 = N-k-1$. Since all those terms are identical except for the two exceptions, in the expression of $\frac{v_{\lambda \lambda} (\alpha)}{v_{\mu \mu} (\alpha)}$ we will get cancellations except for $j =r$ and $i=1,..., N-k-1$ or for $i=N-k$ and $j=1...r$ (note that for $\mu$, when $i=N-k$ $j$ stops at $r-1$). In all other places, the partition constituents and their conjugates are identical and the ratio cancels to identity. We get:
\begin{equation}
\frac{v_{\lambda \lambda} (\alpha)}{v_{\mu^{A_1} \mu^{A_1}} (\alpha)} = \prod_{i=1}^{N-k-1} \frac{\mu^{A_1'}_r - i + 1+ \alpha(\mu^{A_1}_i -r)}{\lambda_r' - i + 1 + \alpha(\lambda_i -r)} \prod_{j=1}^{r-1} \frac{\mu^{A_1'}_j - (N-k) + 1+ \alpha(\mu^{A_1}_{N-k} -j)}{\lambda_j' - (N-k) + 1 + \alpha(\lambda_{N-k} -j)} \frac{1}{\lambda_r' - (N-k) + 1 + \alpha(\lambda_{N-k} - r)}
\end{equation}
Upon massaging, we get:
\begin{equation}
\frac{1}{\lambda_r' - (N-k) + 1 + \alpha(\lambda_{N-k} - r)} =1
\end{equation}
\begin{equation}
\prod_{j=1}^{r-1} \frac{\mu^{A_1'}_j - (N-k) + 1+ \alpha(\mu^{A_1}_{N-k} -j)}{\lambda_j' - (N-k) + 1 + \alpha(\lambda_{N-k} -j)} = \frac{1+ \alpha(r-2)}{1+ \alpha(r-1)}\frac{1+ \alpha(r-3)}{1+ \alpha(r-2)} ... \frac{1+\alpha}{1+ 2 \alpha} \frac{1}{1+\alpha} = \frac{1}{1+ \alpha(r-1)}
\end{equation}
hence
\begin{equation}
\frac{v_{\lambda \lambda} (\alpha)}{v_{\mu^{A_1} \mu^{A_1}} (\alpha)} = \prod_{i=1}^{N-k-1} \frac{\mu^{A_1'}_r - i + 1+ \alpha(\mu^{A_1}_i -r)}{\lambda_r' - i + 1 + \alpha(\lambda_i -r)}\frac{1}{1+ \alpha(r-1)}
\end{equation}
Now for the $B$'s:
\begin{equation}
\frac{\prod_{(i,j) \in \lambda} B^\lambda_{\lambda, \mu^{A_1}}(i,j)}{\prod_{(i,j) \in \mu^{A_1}} B^{\mu^{A_1}}_{\lambda, \mu^{A_1}}(i,j)}= \frac{\prod_{(i,j) \in \lambda, \; j\ne r} B^\lambda_{\lambda, \mu^{A_1}}(i,j)}{\prod_{(i,j) \in \mu^{A_1}, \; j\ne r } B^{\mu^{A_1}}_{\lambda, \mu^{A_1}}(i,j)} * \frac{\prod_{(i,r) \in \lambda} B^\lambda_{\lambda, \mu^{A_1}}(i,r)}{\prod_{(i,r) \in \mu^{A_1}, \; i \ne N-k } B^{\mu^{A_1}}_{\lambda, \mu^{A_1}}(i,r)}
\end{equation}
\noindent where in the last ratio, the differentiation has been made $i\ne N-k$ because $\mu^{A_1}_{N-k}=r-1$ and the point $(i,j) = (N-k,r)$ hence does not belong to the Young tableau of the partition $\mu^{A_1}$. We then have
\begin{equation}
\frac{\prod_{(i,j) \in \lambda, \; j\ne r} B^\lambda_{\lambda, \mu^{A_1}}(i,j)}{\prod_{(i,j) \in \mu^{A_1}, \; j\ne r } B^{\mu^{A_1}}_{\lambda, \mu^{A_1}}(i,j)} =  \frac{\prod_{(i,j) \in \lambda, \; i\ne N-k, \;  j\ne r} B^\lambda_{\lambda, \mu^{A_1}}(i,j)}{\prod_{(i,j) \in \mu^{A_1}, \;i\ne N-k, \; j\ne r } B^{\mu^{A_1}}_{\lambda, \mu^{A_1}}(i,j)} \prod_{j=1}^{r-1} \frac{B^\lambda_{\lambda, \mu^{A_1}}(N-k,j)}{ B^{\mu^{A_1}}_{\lambda, \mu^{A_1}}(N-k,j)}= \prod_{j=1}^{r-1} \frac{B^\lambda_{\lambda, \mu^{A_1}}(N-k,j)}{ B^{\mu^{A_1}}_{\lambda, \mu^{A_1}}(N-k,j)}
\end{equation}
Where the first product simplifies because for $i\ne N-k$ and $j \ne r$ the components of the two partitions and their conjugates are identical. Because for $j=1,...,r-1$ the components of the conjugate partitions $\lambda_j'$ and $\mu^{A_1'}_j$ are equal we have:
\begin{equation}
\prod_{j=1}^{r-1} \frac{B^\lambda_{\lambda, \mu^{A_1}}(N-k,j)}{ B^{\mu^{A_1}}_{\lambda, \mu^{A_1}}(N-k,j)} = \prod_{j=1}^{r-1} \frac{\lambda_j' - (N-k) + \alpha(\lambda_{N-k} - j +1)}{\mu^{A_1'} - (N-k) + \alpha (\mu^{A_1}_{N-k} - j+1)} = \prod_{j=1}^{r-1}\frac{r+1-j}{r-j} = r
\end{equation}
and we have solved half the products that make up the ratio of the $B$'s.

We also must refine, since $\lambda_r' \ne \mu^{A_1'}_r$:
\begin{eqnarray}
&\frac{\prod_{(i,r) \in \lambda} B^\lambda_{\lambda, \mu^{A_1}}(i,r)}{\prod_{(i,r) \in \mu^{A_1}, \; i \ne N-k } B^{\mu^{A_1}}_{\lambda, \mu^{A_1}}(i,r)} = \prod_{i=1}^{N-k-1} \frac{\lambda_r' - i +1 + \alpha(\lambda_i -r)}{\mu^{A_1'}_r - i +1 + \alpha (\mu^{A_1}_i -r)} (\lambda_r' - (N-k)+ 1 + \alpha(\lambda_{N-k}-r)) = \nonumber \\ &= \prod_{i=1}^{N-k-1} \frac{\lambda_r' - i +1 + \alpha(\lambda_i -r)}{\mu^{A_1'}_r - i +1 + \alpha (\mu^{A_1}_i -r)}
\end{eqnarray}
To get to the equation:
\begin{equation}
\frac{\prod_{(i,j) \in \lambda} B^\lambda_{\lambda, \mu^{A_1}}(i,j)}{\prod_{(i,j) \in \mu^{A_1}} B^{\mu^{A_1}}_{\lambda, \mu^{A_1}}(i,j)} = r\cdot \prod_{i=1}^{N-k-1} \frac{\lambda_r' - i +1 + \alpha(\lambda_i -r)}{\mu^{A_1'}_r - i +1 + \alpha (\mu^{A_1}_i -r)}
\end{equation}

We finally reach the solution:
\begin{equation}
A_1= \beta_{\lambda \mu^{A_1}} = \prod_{i=1}^{N-k-1} \frac{\mu^{A_1'}_r - i + 1+ \alpha(\mu^{A_1}_i -r)}{\lambda_r' - i + 1 + \alpha(\lambda_i -r)}\frac{1}{1+ \alpha(r-1)}
 r\cdot \prod_{i=1}^{N-k-1} \frac{\lambda_r' - i +1 + \alpha(\lambda_i -r)}{\mu^{A_1'}_r - i +1 + \alpha (\mu^{A_1}_i -r)} = \frac{r}{1+ \alpha(r-1)}
\end{equation}
which is $-1/\nu$ as mentioned in the text.  Notice the massive number of cancelations that occur upon multiplying the $B$'s with the $v$'s, which could be done from the very beginning in the formula but which would then obscure the meaning of the two terms.  The reader is again warned that in the literature it is more common to use a different Jack normalization if one is interested in combinatoric formulas for which the other normalization is more suitable.

\section{Coefficient $B_1$}

The partition $\mu^{B_1}$ defined previously, has the following properties: $mu^{B_1}_i = \lambda_i$, $\forall$ $i=1,..., N-k-1$ and $\mu^{B_1}_{N-k} =\lambda_{N-k} -2 = r-2$. The conjugate partition has the following properties: $\mu^{B_1'}_i = \lambda_i'$ $\forall$ $i=1,...,r-2$ and $i=r+1,...,\frac{r}{k}(N-k)$ and $\mu^{B_1'}_{r-1}=\mu^{B_1'}_{r} = \lambda_{r-1}'-1 = \lambda_r'-1 = N-k-1$
We then have:
\begin{eqnarray}
&\frac{v_{\mu^{B_1} \mu^{B_1}}}{v_{\lambda \lambda}} = \frac{\prod_{i=1}^{N-k-1} \prod_{j=1}^{\mu^{B_1}_i} h^{\mu^{B_1}}_*(i,j)}{\prod_{i=1}^{N-k-1} \prod_{j=1}^{\lambda_i} h^\lambda_*(i,j)} \frac{\prod_{j=1}^{\mu^{B_1}_{N-k}}h^{\mu^{B_1}}_*(N-k, j)}{\prod_{j=1}^{\lambda_{N-k}} h^\lambda_*(N-k,j)} = \nonumber \\ & = \prod_{i=1}^{N-k-1}\frac{ h^{\mu^{B_1}}_*(i,r-1) h^{\mu^{B_1}}_*(i,r)}{h^\lambda_*(i, r-1) h^\lambda_*(i,r)} \cdot \prod_{j=1}^{r-2} \frac{ h^{\mu^{B_1}}_*(N-k, j)}{h^\lambda_*(N-k,j)} \frac{1}{h^\lambda_*(N-k,r-1) h^\lambda_*(N-k,r)} = \nonumber \\  & = \prod_{i=1}^{N-k-1}\frac{ h^{\mu^{B_1}}_*(i,r-1) h^{\mu^{B_1}}_*(i,r)}{h^\lambda_*(i, r-1) h^\lambda_*(i,r)} \cdot \prod_{j=1}^{r-2} \frac{1+ \alpha(r-2-j)}{1+ \alpha(r-j)} \frac{1}{1+\alpha} =\nonumber \\  & = \prod_{i=1}^{N-k-1}\frac{ h^{\mu^{B_1}}_*(i,r-1) h^{\mu^{B_1}}_*(i,r)}{h^\lambda_*(i, r-1) h^\lambda_*(i,r)} \frac{1}{(1+\alpha(r-1))(1+ \alpha(r-2))}
\end{eqnarray}
Now for the $B$'s:
\begin{eqnarray}
&\frac{\prod_{(i,j) \in \lambda} B^\lambda_{\lambda, \mu^{B_1}}(i,j)}{\prod_{(i,j) \in \mu^{B_1}} B^{\mu^{B_1}}_{\lambda, \mu^{B_1}}(i,j)}= \frac{\prod_{i=1}^{N-k-1} \prod_{j=1}^{\lambda_i} B^\lambda_{\lambda, \mu^{B_1}}(i,j)}{\prod_{i=1}^{N-k-1} \prod_{j=1}^{\mu^{B_1}_i} B^{\mu^{B_1}}_{\lambda, \mu^{B_1}}(i,j)} \frac{\prod_{j=1}^{\lambda_{N-k}} B^\lambda(N-k,j)}{\prod_{j=1}^{\mu^{B_1}_{N-k}} B^{\mu^{B_1}}(N-k,j)} = \noindent \\ & = \prod_{i=1}^{N-k-1} \frac{B^\lambda(i,r-1) B^\lambda(i,r)}{B^{\mu^{B_1}}(i,r-1) B^{\mu^{B_1}}(i,r)} \prod_{j=1}^{r-2}\frac{B^\lambda(N-k,j)}{B^{\mu^{B_1}}(N-k,j)} B^\lambda(N-k, r-1) B^\lambda(N-k,r) = \nonumber \\ & = \prod_{i=1}^{N-k-1} \frac{h^\lambda_*(i, r-1) h^\lambda_*(i,r)}{ h^{\mu^{B_1}}_*(i,r-1) h^{\mu^{B_1}}_*(i,r)} \prod_{j=1}^{r-2}\frac{h_{\lambda}^*(N-k,j)}{h_{\mu^{B_1}}^* (N-k,j)} h^\lambda_*(N-k,r-1) h^\lambda_*(N-k,r) = \nonumber \\ & = \prod_{i=1}^{N-k-1} \frac{h^\lambda_*(i, r-1) h^\lambda_*(i,r)}{ h^{\mu^{B_1}}_*(i,r-1) h^{\mu^{B_1}}_*(i,r)} \prod_{j=1}^{r-2} \frac{r-j+1}{r-j-1} (1+\alpha) = \prod_{i=1}^{N-k-1} \frac{h^\lambda_*(i, r-1) h^\lambda_*(i,r)}{ h^{\mu^{B_1}}_*(i,r-1) h^{\mu^{B_1}}_*(i,r)} \frac{r(r+1)}{2} (1+\alpha)
\end{eqnarray}

By multiplying the $v$'s and the $B$'s, and canceling the common factor (which again could have been canceled earlier) one gets:

\begin{equation}
B_1 = \frac{r(r-1)}{2} \frac{1+\alpha}{(1+ \alpha(r-1))(1+ \alpha(r-2))}
\end{equation}

\section{Some Jack Polynomial Identities}
\label{app:vanish}

Lassalle\cite{Lassalle} found
the following identity for Jack polynomials (the normalization in Lassale's paper is different than the normalization we use, so this formula has been modified appropriately)
\begin{equation}
\sum_i \frac{\partial}{\partial z_i} J^\alpha_{\{\lambda\}} =
\sum_{m} A_{\{\lambda\},\{\lambda_{(m)}\}}
J_{\{\lambda_{(m)}\}}^\alpha
\end{equation}
\noindent where the coefficient reads:
\begin{eqnarray}
& A_{\{\lambda\},\{\lambda_{(m)}\}} = \frac{1}{\alpha}
 \left( \prod_{j=m+1}^{l_{\lambda}} \frac{
  \alpha (\lambda_m - \lambda_j)+j-m-1}{  \alpha (\lambda_m -
 \lambda_j)+ j-m} \right)\left( \prod_{j=1}^{\lambda_m -1}
 \frac{\lambda_j^, -m+1 + \alpha(\lambda_m -j -1)}{\lambda_j^,-m+1 + \alpha(\lambda_m
 -j)} \right) \times \nonumber \\ & \times (l_{\lambda} - m + \alpha \lambda_m)(N - m+1 + \alpha (\lambda_m-1))
\label{lassalle}
\end{eqnarray}
\noindent where $\lambda_{(m)}$ is the partition (elements of partitions are denoted by $\lambda_i$, but $\lambda_{(m)}$ is a full partition) where you remove $1$
from the row $\lambda_m$ in the partition $\lambda$, and where
$l_{\lambda}$ is the length of the partition $\lambda$, and $\lambda'$ is the partition conjugate to $\lambda$ - this means write the partition $\lambda =(\lambda_1, \lambda_2,\ldots,\lambda_N)$ as a Young diagram with the $i$'th row of length $\lambda_i$, then transpose (as in matrix transpose) this to get $\lambda'$.  As another example,  $\lambda_{(m)}$ is the partition where $1$ is subtracted from $\lambda_m$ in $\lambda$ : if $\lambda= (4,4,2,2)$ then $\lambda_{(1)}$ doesnt exist because the partition one would obtain is then $(3,4,2,2)$ which doesnt satisfy the rule that the partition must be made of decreasing integers. Then $\lambda_{(2)}=(4,3,2,2)$, $\lambda_{(3)}$ doesnt exist, and $\lambda_{(4)} = (4,4,2,1)$.

It is now easy to prove that
\begin{equation}
\sum_i \frac{\partial}{\partial z_i} J_{0k0^{r-1}k0^{r-1}k\ldots k0^{r-1}k} = J_{1k-10^{r-1}k0^{r-1}k\ldots k0^{r-1}k} \label{firstorder}
\end{equation}
First, lets translate everything in partition language:
\begin{equation}
[0k0^{r-1}k0^{r-1}k\ldots k0^{r-1}k]\rightarrow \lambda \left(\underbrace{ \frac{r}{k}(N-k)+1}_{k \text{ times}}, \underbrace{\frac{r}{k}(N-k)+1-r}_{k \text{ times}}, \ldots, \underbrace{2r+1}_{k \text{ times}},\underbrace{ r+1}_{k \text{ times}}, \underbrace{1}_{k \text{ times}} \right)
\end{equation}
First, let me find the coefficient of $J_{1(k-1)0^{r-1}k0^{r-1}k\ldots k0^{r-1}k}$ in $\sum_i \frac{\partial}{\partial z_i} J_{0k0^{r-1}k0^{r-1}k\ldots k0^{r-1}k}$.

We have $l_{\lambda} = N$. In Lassale's notation, ${1k-10^{r-1}k0^{r-1}k\ldots k0^{r-1}k}$ corresponds to $\lambda_{(N)}$, i.e. the partition where $1$ was subtracted from $\lambda_N=1$ so, the first two products in Eq.(\ref{lassalle}) do not contribute so:
\begin{equation}
A_{\{\lambda\},\{\lambda_{(N)}\}} =\frac{1}{\alpha} (N - N + \alpha \lambda_N)(N - N+1 + \alpha (\lambda_N-1)) = 1
\end{equation}
so we proved that the coefficient is $1$, as we wanted. To prove Eq.(\ref{firstorder}), we also need to prove that all other contributions vanish. I.e. we have a bunch of other $\lambda_{(m)}$'s which we can write as:
\begin{equation}
 \lambda_{(N- p k)}=  \left(\underbrace{ \frac{r}{k}(N-k)+1}_{k \text{ times}}, \underbrace{\frac{r}{k}(N-k)+1-r}_{k \text{ times}}, \ldots,\underbrace{(p+1) r}_{k \text{ times}},\underbrace{p r+1}_{k-1 \text{ times}}, p r , \underbrace{(p-1) r}_{k \text{ times}},\ldots,\underbrace{2r+1}_{k \text{ times}},\underbrace{ r+1}_{k \text{ times}}, \underbrace{1}_{k \text{ times}} \right)
\end{equation} where $p$ is an integer in the interval $[0,\ldots, \frac{N}{k}-1]$. In the original partition, $\lambda$ the component $\lambda_{N- pk} = p r +1$ It is now easy to see that the coefficient $A_{\{\lambda\},\{\lambda_{(N- p k )}\}} =0$ for any $p$. The reason is the the second product on the RHS in the first row of Eq.(\ref{lassalle}) vanishes. The key is the numerator:
\begin{equation}
\prod_{j=1}^{\lambda_m -1}
 (\lambda_j^, -m+1 + \alpha(\lambda_m -j -1))=
\prod_{j=1}^{\lambda_{N- p k} -1}
 (\lambda_j^, -(N- p k) +1 + \alpha(\lambda_{N-p k} -j -1)) = \prod_{j=1}^{p r}
 (\lambda_j^, -(N- p k) +1 + \alpha(p r -j))
\end{equation}
\noindent We now want to look at the term in the product that has $j_0=(p-1) r+1$. One can immediately see that $\lambda^,_{j_0} = N- (p-1) k $. The product above then becomes:
\begin{equation}
N - (p-1) k - (N- p k) + 1 + \alpha( p r - (p-1) r-1) = k+1 + \alpha ( r-1)=0
\end{equation}
\noindent where we have used the fact that we are looking at Jacks with $\alpha = - (k+1)/(r-1)$. As such we have proved Eq.(\ref{firstorder}). But we also know that $J_{0k0^{r-1}k0^{r-1}k\ldots k0^{r-1}k} = \prod_{i=1}^N z_i J_{k0^{r-1}k0^{r-1}k\ldots k0^{r-1}k}$ and hence we obtain:
\begin{eqnarray}
\sum_i \frac{\partial}{\partial z_i} J_{0k0^{r-1}k0^{r-1}k\ldots k0^{r-1}k} &=& (J_{1(k-1)0^{r-1}k0^{r-1}k\ldots k0^{r-1}k} ) =  \sum_i \frac{\partial}{\partial z_i} \prod_{j=1}^N z_j J_{k0^{r-1}k0^{r-1}k\ldots k0^{r-1}k} \\ &=& \sum_i \frac{1}{z_i} J_{0k0^{r-1}k0^{r-1}k\ldots k0^{r-1}k}
\end{eqnarray}
\noindent where we have used the fact that $\sum_i \frac{\partial}{\partial z_i} J_{k0^{r-1}k0^{r-1}k\ldots k0^{r-1}k}=0$ by virtue that $J_{k0^{r-1}k0^{r-1}k\ldots k0^{r-1}k}$ is a Highest Weight translationaly invariant ground state\cite{Bernevig1}.

\section{Further Jack Identities}
\label{app:onefurther}

Proposition 5.1 of Ref.~\oncite{Stanley} says that:
\begin{equation}
J^\alpha_{\lambda}(z_1,...,z_N) = \prod_i z_i J^\alpha_{\lambda -I} (z_1,...,z_N)
\end{equation}
where $\lambda- I= (\lambda_1- 1, \lambda_2-1,..., \lambda_n -1)$ where $n$ is the length of the partition. This of course supposes that $\lambda_n>0$ which means that the zeroth orbital, in occupation number language must be zero. This proves our equations \ref{eq:efirst}, \ref{eq:esecond} and \ref{eq:ethird}.

We then have (the sum over the particles $i$ goes from $k+1$ to $N$, but the numbers of particles is explicit in the occupation number of any partition)
\begin{eqnarray}
&\sum_i \frac{\partial}{\partial z_i} J^\alpha_{0^{r}k0^{r-1}k0^{r-1}...k0^{r-1}k}  = \sum_i \frac{\partial}{\partial z_i}  \prod_{i} z_i^{r} J^\alpha_{k0^{r-1}k0^{r-1}...k0^{r-1}k} =  \nonumber \\ & = r \sum_i \frac{1}{z_i} \prod_i z_i^{r} J^\alpha_{k0^{r-1}k0^{r-1}...k0^{r-1}k} + \prod_i z_i^{r} \sum_i \frac{\partial}{\partial z_i} J^\alpha_{k0^{r-1}k0^{r-1}...k0^{r-1}k} = r  \prod_i z_i^{r-1}\sum_i \frac{1}{z_i}  J^\alpha_{0k0^{r-1}k0^{r-1}...k0^{r-1}k}
\end{eqnarray}
Now by equations D9, D10, and D3 in our paper we get

\begin{equation}
 \prod_i z_i^{r-1}\sum_i \frac{1}{z_i}  J^\alpha_{0k0^{r-1}k0^{r-1}...k0^{r-1}k} =  \prod_i z_i^{r-1}  J^\alpha_{1k-10^{r-1}k0^{r-1}...k0^{r-1}k} = J^\alpha_{0^{r-1}1k-10^{r-1}k0^{r-1}...k0^{r-1}k}
\end{equation}
so
\begin{equation}
\sum_i \frac{\partial}{\partial z_i} J^\alpha_{0^{r}k0^{r-1}k0^{r-1}...k0^{r-1}k}  = r J^\alpha_{0^{r-1}1k-10^{r-1}k0^{r-1}...k0^{r-1}k}
\end{equation}


\begin{thebibliography}{99}
\addcontentsline{toc}{section}{References}

\bibitem{Prange}  For a classic review of quantum Hall physics,
see R. Prange and S. M. Girvin eds, {\it The Quantum Hall Effect},
Springer-Verlag, NY (1987).


\bibitem{RMP} For a recent general review see C. Nayak, S. H. Simon, A. Stern, M. Freedman, S. DasSarma, {\it Non-Abelian Anyons and Topological Quantum Computation}, Rev. Mod. Phys.  {\bf 80}, 1083, (2008);  http://www.arxiv.org/abs/0707.1889  and references therein.

\bibitem{YellowBook} P. Di Francesco,  P. Mathieu, and D. S´en´echal, {\it Conformal FieldTheory}, Springer, NewYork, (1997).


\bibitem{MooreRead}  G. Moore and N. Read,  {\it Nonabelions in the fractional quantum hall effect} Nucl.Phys. {\bf B360} (1991) 362.

\bibitem{ReadRezayi} N. Read and E. Rezayi, {\it Beyond paired quantum Hall states: Parafermions and incompressible states in the first excited Landau level}, Phys. Rev. B {\bf 59}, 8084 (1999)


\bibitem{Bernevig1}  B. A. Bernevig and F. D. M. Haldane {\it
 Model Fractional Quantum Hall States and Jack Polynomials}, Phys. Rev. Lett. {\bf 100}, 246802 (2008).

\bibitem{Bernevig2} B. A. Bernevig and F. D. M. Haldane, {\it Generalized clustering conditions of Jack polynomials at negative Jack parameter},  Phys. Rev. B {\bf 77}, 184502 (2008).

\bibitem{Bernevig3}B. A. Bernevig and F. D. M. Haldane, {\it
Properties of Non-Abelian Fractional Quantum Hall States at Filling $\nu=k/r$}, . arXiv:0803.2882


\bibitem{Jack} H. Jack, "A class of symmetric polynomials with a parameter", Proc. Roy. Soc. Edinburgh Sect. A, {\bf 69}, 1-18, 1970.


\bibitem{Green} D. Green, PhD. Thesis (2002); available online at arXiv:cond-mat/0202455.

\bibitem{WenWu} 	
	X.-G. Wen and Y.-S. Wu,
	{\it Chiral operator product algebra hidden in certain fractional quantum Hall wave functions}, Nuc. Phys. {\bf B419}, 45 (1994)


\bibitem{Gaffnian} S. H. Simon, E. H. Rezayi, N. R. Cooper, and I. Berdnikov,
{\it Construction of a paired wave function for spinless electrons at
filling fraction $\nu$=2/5}, Phys. Rev. B {\bf 75}, 075317 (2007).


\bibitem{Jimbo1} B. Feigin, M. Jimbo, T. Miwa and E. Mukhin, {\it A differential ideal of symmetric polynomials spanned by Jack polynomials at $\beta = -(r-1)/(k+1)$}, International Mathematics Research Notices 1223 (2002); arXiv:math/0112127.

\bibitem{Jimbo2} B. Feigin, M. Jimbo, T. Miwa and E. Mukhin, {\it Symmetric polynomials vanishing on the shifted diagonals and Macdonald polynomials}, International Mathematics Research Notices 1015 (2003); arXiv:math/0209042.

\bibitem{jolicoeur} M.V. Milovanovic, Th. Jolicœur, I. Vidanovic, {\it Healing non-unitary conformal field theories and related fractional quantum Hall states}, arXiv:0902.1719.



\bibitem{Bernevig4} B. A. Bernevig, F.D.M. Haldane, {\it Clustering Properties and Model Wavefunctions for Non-Abelian Fractional Quantum Hall Quasielectrons},  Phys. Rev. Lett. {\bf{102}}, 066802 (2009);  arXiv:0810.2366.

\bibitem{Bernevig5} N. Regnault, B. Andrei Bernevig, F.D.M. Haldane, {\it Topological Entanglement and Clustering of Jain Hierarchy States},  arXiv:0901.2121.


\bibitem{HaldaneAPS}  F.D.M. Haldane, , Bull. Am. Phys. Soc. 51, 633 (2006).


\bibitem{Stanley} R. P. Stanley {\it Some Combinatorial Properties of Jack Symmetric Functions}, Adv. Math. {\bf 77}, 76 (1989).


\bibitem{Demmel} J. Demmel and P. Koev, {\it Accurate and efficient evaluation of Schur and Jack functions}, Math. Comp., {\bf 75},  223, 2006.


\bibitem{MacDonald} I. G. Macdonald, Symmetric functions and Hall polynomials, Second ed., Oxford University Press, New York, 1995.

\bibitem{Lassalle}  M. Lassalle,  {\it Coefficients binomiaux generalises et polynomes de Macdonald}, J. Funct. Anal. 158, 289 (1998).


\bibitem{ReadViscocity} N. Read, {\it Non-Abelian adiabatic statistics and Hall viscosity in quantum Hall states and $p_x+ip_y$ paired superfluids}, arXiv:0807.3107.


\bibitem{ReadEdge} N. Read,    {\it  Conformal invariance of chiral edge theories}, arXiv:0711.0543



\bibitem{Zamparafermion}
A.~Zamolodchikov and V.~Fateev,
{\it Nonlocal (Parafermion) Currents in Two-Dimensional Conformal Quantum
  Field Theory and Self-Dual Critical Points in $Z_N$-Symmetrical Statistical
  Systems},
 Sov. Phys. JETP, {\bf 62} (1985) 215--225.







\bibitem{Jacob} P. Jacob  and P. Mathieu  {\it The $Z_k^{(su(2),3/2)}$ parafermions}, Physics letters B., 627, 224 (2005)

\bibitem{Ravanini} F. Ravanini, {\it On the possibility of  $Z_N$ exotic supersymmetry in two dimensional conformal field theory}, Int. J. Mod. Phys. {\bf A7}  4949 (1992).

\bibitem{SimonProjection} S. H. Simon, E. H. Rezayi, and N. R. Cooper, {\it Generalized quantum Hall projection Hamiltonians}, Phys. Rev. B {\bf 75},
075318 (2007)


\bibitem{Bergholtz} E. J. Bergholtz, T. H. Hansson, M. Hermanns, and A. Karlhede , {\it Microscopic Theory of the Quantum Hall Hierarchy}, Phys. Rev. Lett. 99, 256803 (2007).

\bibitem{Bergholtz2}   E.J. Bergholtz, T.H. Hansson, M. Hermanns, A. Karlhede, S. Viefers, {\it Hierarchy wave functions--from conformal correlators to Tao-Thouless states}, Phys. Rev. B 77, 165325 (2008). 

\bibitem{Seidel} A. Seidel and D.-H. Lee, {\it Abelian and Non-Abelian Hall Liquids and Charge-Density Wave: Quantum Number Fractionalization in One and Two Dimensions}, Phys. Rev. Lett. 97, 056804 (2006).


\bibitem{ArdonneTorus} E. Ardonne, E. J. Bergholtz, J. Kailasvuori, E.~Wikberg, {\it Degeneracy of non-abelian quantum Hall states on the torus: domain walls and conformal field theory}, J. Stat. Mech. P04016 (2008)


\bibitem{ArdonneTorus2} E. Ardonne, {\it Domain walls, fusion rules and conformal field theory in the quantum Hall regime}, arXiv:0809.0389.

\bibitem{Gurarie} V. Gurarie and C. Nayak, {\it A Plasma Analogy and Berry Matrices for Non-Abelian Quantum Hall States}, Nucl. Phys. {\bf B506}, 685 (1997).

\bibitem{Nayak} C. Nayak and F. Wilczek, {\it $2n$ Quasihole States Realize $2^{n-1}$-Dimensional Spinor Braiding Statistics in Paired Quantum Hall States}, Nucl. Phys. {\bf B479}, 529 (1996)


\bibitem{SchoutensReview} A good review of ${\cal W}$-symmetry in conformal field theory is given by
P. Bouwknegt and K. Schoutens, {\it W symmetry in conformal field theory}, Phys.Rept. {\bf 223} (1993) 183.


\bibitem{Fateev1} V. A. Fateev and S. L. Lykyanov, {\it The Models of Two-Dimensional Conformal Quantum Field Theory with $Z_n$ Symmetry} Int. J. Mod.
Phys. {\bf A3} (1988) 507.

\bibitem{Fateev2}  V. A. Fateev and S. L. Lykyanov, {\it Additional Symmetries and Exactly Solvable Models in Two-Dimensional Conformal Field Theory}, Sov. Sci. Rev. A. Phys.
 {\bf 15}/{\bf 2}, parts I,II,III
(1990).






\bibitem{Bais} F. A. Bais, P. Bouwknegt, K. Schoutens, and M.
Surridge, {\it Extensions of the Virasoro algebra constructed from Kac-Moody algebras using higher order Casimir Invariants}, Nucl. Phys. {\bf B304} (1988) 348; {\it ibid} p. 371.


\bibitem{endnotealpha} We follow the convention of Ref.
\oncite{Fateev1} for these variables.  The definitions of Ref.
\oncite{Fateev2} differ by factors of 2.

\bibitem{Mizoguchi} S. Mizoguchi, {\it Non-Unitarity Theorem for the A type $W_n$ Algebra}, Phys. Lett. {\bf B231}, 112 (1989).

\bibitem{Dunning} C. Dunning, {\it Massless Flows Between Minimal $W$ Models}, Phys. Lett. {\bf B537}, 297 (2002).


\bibitem{Schweigert} J. Froehlich, B. Pedrini, C. Schweigert, and J. Walcher, {\it Universality in Quantum Hall Systems:
Coset Construction of Incompressible States}, 	J. Stat. Phys. {\bf 103}, 527  (2004).



\bibitem{Calogero} See for example, P. J. Forrester, {\it Jack Polynomials and the Multi-component Calogero-Sutherland model}, Int. J. Mod. Phys. {\bf B10} 427 (1996), and therein.


\bibitem{Simon3} S. H. Simon, E. H. Rezayi, and N. R. Cooper
, {\it Pseudopotentials for multiparticle interactions in the quantum Hall regime}, Phys. Rev. B 75, 195306 (2007)

\end{thebibliography}
\end{document}